%% file: main.tex
\documentclass{amsart}
\usepackage{amsmath}
\usepackage{amssymb}
\usepackage{amsaddr}
\usepackage{amsthm}
\usepackage{stmaryrd}
\usepackage{siunitx}
\usepackage[numbers]{natbib}
\usepackage{tikz}
\usetikzlibrary{arrows,shapes,trees,patterns} 
\usepackage{subfig}
\usepackage{color}
\usepackage{array}
\usepackage{graphicx}
\usepackage{stmaryrd}
\usepackage{pgfplots}
\usepackage{bbm}
\usepackage{aicescover}

\newcommand{\dx}{\,{\rm d}\mathbf{x}}
\newcommand{\ds}{\,{\rm d}\sigma}

\newcommand{\jump}[1]{\left\llbracket #1\right\rrbracket}
\newcommand{\vint}[1]{\left(#1\right)_{\mathcal{T}_h}}
\newcommand{\eint}[1]{\left<#1\right>_{\partial\mathcal{T}_h}}
\newcommand{\beint}[1]{\left<#1\right>_{\Gamma^b_h}}
\newcommand{\eints}[1]{\left<#1\right>_{\Gamma_h}}

\usetikzlibrary{patterns}
\usetikzlibrary{calc}
\newcommand*{\boxcolor}{red}

\pgfplotsset{
  label style={anchor=near ticklabel},
  xlabel style={yshift=0.5em},
  ylabel style={yshift=-1em},
  tick label style={font=\scriptsize},
  label style={font=\scriptsize},
  legend style={font=\scriptsize},
  cycle list={{black, mark=x}, {red,mark=triangle},{blue, mark=square}, {green,mark=+},{magenta, mark=diamond}, {orange,mark=x}}}

\makeatletter
\def\maketag@@@#1{\hbox{\m@th\normalfont\normalsize#1}} 
\newcommand{\miniscule}{\@setfontsize\miniscule{4}{5}}
\renewcommand{\boxed}[1]{\textcolor{\boxcolor}{%
\tikz[baseline={([yshift=-1ex]current bounding box.center)}] \node [rectangle, minimum width=1ex,rounded corners,draw] {\normalcolor\m@th$\displaystyle#1$};}}
\newcommand\resetstackedplots{

\pgfplots@stacked@isfirstplottrue
\makeatother
\addplot [forget plot,draw=none] coordinates{(1,0) (2,0) (3,0)};}
\makeatother

\title[A Comparison of HDG and DG for Target-Based $hp$-Adaptation]{A Comparison of Hybridized and Standard DG Methods for Target-Based $hp$-Adaptive Simulation of Compressible Flow}

\author[Woopen, Balan, May, Sch\"utz]{Michael Woopen, Aravind Balan, Georg May}

\address[aices]{Aachen Institute for Advanced Study in Computational Engineering Science, RWTH Aachen University, Aachen, Germany}

\author{Jochen Sch\"utz}

\address[igpm]{Institut f\"ur Geometrie und Praktische Mathematik,                     RWTH Aachen University, Aachen, Germany}

\aicescoverauthor{Michael Woopen, Aravind Balan, Georg May, and Jochen Sch\"utz}

\begin{document}

\aicescoverpage

\maketitle

\begin{abstract}
We present a comparison between hybridized and non-hybridized discontinuous Galerkin methods in the context of target-based $hp$-adaptation for compressible flow problems. The aim is to provide a critical assessment of the computational efficiency of hybridized DG methods.

Hybridization of finite element discretizations has the main advantage, that the resulting set of algebraic equations has globally coupled degrees of freedom only on the skeleton of the computational mesh. Consequently, solving for these degrees of freedom involves the solution of a potentially much smaller system. This not only reduces storage requirements, but also allows for a faster solution with iterative solvers. Using a discrete-adjoint approach, sensitivities with respect to output functionals are computed to drive the adaptation. From the error distribution given by the adjoint-based error estimator, $h$- or $p$-refinement is chosen based on the smoothness of the solution which can be quantified by properly-chosen smoothness indicators. 

Numerical results are shown for subsonic, transonic, and supersonic flow around the NACA0012 airfoil. $hp$-adaptation proves to be superior to pure $h$-adaptation if discontinuous or singular flow features are involved. In all cases, a higher polynomial degree turns out to be beneficial. We show that for polynomial degree of approximation $p=2$ and higher, and for a broad range of test cases, HDG performs better than DG in terms of runtime and memory requirements.
\end{abstract}



\input{introduction}

\input{physics}

\input{discretization}

\input{errorestimation}

\input{results}

\input{conclusion}

\section*{Acknowledgments}

Financial support from the Deutsche Forschungsgemeinschaft (German Research Association) through grant GSC 111 is gratefully acknowledged.

\bibliographystyle{plainnat}

\end{document}

%% file: introduction.tex

\section{Introduction}
During the last years, discontinuous Galerkin (DG)  methods (see, e.g., \cite{Bassi1997,DG2, ABCM}) have become increasingly popular.
This is indisputable due to their advantages --- high-order accuracy on unstructured meshes, a variational setting, and local conservation, just to name a few.

However, the use of discontinuous function spaces is at the same time the reason for a major disadvantage: unlike in continuous Galerkin (CG) methods, degrees of freedom are not shared between elements. As a consequence, the number of unknowns is substantially higher compared to a CG discretization. Especially for implicit time discretization this imposes large memory requirements, and potentially leads to increased time-to-solution.

In order to avoid these disadvantages, a technique called hybridization may be utilized (see \cite{fraeijs1965displacement, cockburn2004characterization, cockburn2009unified, nguyen2009implicitlin, NPC09, peraire2010hybridizable, AA-HARP:13}), resulting in hybridized discontinuous Galerkin (HDG) methods. Here, the globally coupled unknowns have support on the mesh skeleton, i.e. the element interfaces, only. This reduces the size of the global system and coincidentally improves the sparsity pattern.

However, aiming at industry applications, e.g.\, turbulent flow around a complete airplane or within an aircraft engine, hybridization alone does most likely not provide a sufficiently successful overall algorithm. In these applications one is usually interested in certain quantities only, for example lift or drag coefficients in aerospace, instead of the solution quality per se. Thus, it might be beneficial to distribute the degrees of freedom within the computational domain in such a way that the solution to the discretized problem is close to optimal with respect to the accuracy of these quantities. To achieve this goal, target-based error control methods have been developed (see \cite{hartmann2002adaptive, venditti2002grid, venditti2003anisotropic, hartmann2006adaptive, fidkowski2011review}). One such method is based on the adjoint solution of the original governing equations with respect to the target functional. In this method, an additional linear system of partial differential equations is solved 
which then gives an estimate on the spatial error distribution contributing to the error in the target functional. This estimate can be used as a criterion for local adaptation. Within the context of low order schemes, mesh refinement is used for adaptation~\cite{venditti2002grid, venditti2003anisotropic}. 
Using DG (or HDG), however, offers the additional possibility of varying the polynomial degree within each element. For smooth solutions, this is more efficient compared to mesh refinement, as it yields exponential convergence. In the context of wave problems, \citet{giorgiani2013hybridizable} showed the benefit of using $p$-adaptation within an HDG-framework. Combining both mesh- and order-refinement results in so-called $hp$-adaptation.

In \cite{schutz201358}, we presented a discretization method for nonlinear convection-diffusion equations. The method is based on a discontinuous Galerkin discretization for convection terms, and a mixed method using H(div) spaces for the diffusive terms. Furthermore, hybridization is used to reduce the number of globally coupled degrees of freedom. Adjoint consistency was shown in \cite{schutz2012adjoint}.
In \cite{woopen2013adjoint, balan2013hybridadjointhp}, we extended our computational framework to include HDG schemes, as well as adjoint-based $h$- and $hp$-adaptation. In the current paper, we compare our HDG method with a standard DG method in the context of $hp$-adaptation for stationary compressible flow, mainly with the aim to assess the efficiency of both methods.

This paper is structured as follows. We  briefly cover the governing equations, namely the compressible Euler and Navier-Stokes equations, in Sec.\,\ref{sec:physics}. After that we introduce our discretization and describe the concept of hybridization in Sec.\,\ref{sec:discretization}. In Sec.\,\ref{sec:errorestimation} we establish the adjoint formulation and show how hybridization can be applied to the dual problem. Then we show its efficiency and robustness with examples from compressible flow, including the subsonic, transonic, and supersonic regime, in Sec.\,\ref{sec:results}. 
Finally, we offer conclusions and outlook on future work in Sec.\,\ref{sec:conclusions}.

%% file: physics.tex
\section{Governing Equations}
\label{sec:physics}



We consider systems of partial differential equations 
\begin{equation}
 \boldsymbol\nabla\cdot\left(\mathbf{f}_c(w)-\mathbf{f}_v(w,\boldsymbol\nabla w)\right)=s\left(w,\boldsymbol\nabla w\right)
\end{equation}
with convective and diffusive fluxes 
\begin{equation}
 \mathbf{f}_c:\mathbb{R}^m\rightarrow \mathbb{R}^{m\times d} \quad \text{and} \quad \mathbf{f}_v:\mathbb{R}^m\times \mathbb{R}^{m\times d}\rightarrow\mathbb{R}^{m\times d}, 
\end{equation}
respectively, and a state-dependent source term 
\begin{equation}
 s:\mathbb{R}^m\times\mathbb{R}^{m\times d}\rightarrow\mathbb{R}^m
\end{equation}
on domain $\Omega \subset \mathbb{R}^d$.
Potentially, some of these quantities could be zero. We denote the spatial dimension by $d$ and the number of conservative variables by $m$.
Boundary conditions can be applied either to the conservative variables $w\in\mathbb{R}^m$ and their gradient $\boldsymbol\nabla w\in\mathbb{R}^{m\times d}$ or directly to the fluxes $\mathbf{f}_c$ and $\mathbf{f}_v$.

\subsection{Two-Dimensional Euler Equations}
The Euler equations are comprised of the inviscid compressible continuity, momentum and energy equations. 
They are given in conservative form as
\begin{equation}
\boldsymbol\nabla\cdot \mathbf{f}_c(w)=0
\end{equation}
with the vector of conserved variables
\begin{equation}
w=(\rho,\rho \mathbf{v}, E)^T
\end{equation}
where $\rho$ is the density, $\bf v$ is the velocity vector $\mathbf{v}:=(v_x,v_y)^T$, and $E$ the total energy.
The convective flux is given by 

\begin{equation}
\mathbf{f}_c=\left(\rho \mathbf{v},p\,\mathbf{Id}+\mathbf{v}\otimes\mathbf{v},\mathbf{v}(E+p)\right)^T.
\end{equation}

Pressure is related to the conservative flow variables $w$ by the equation of state
\begin{equation}
p=(\gamma-1)\left(E-\frac12\rho\mathbf{v}\cdot\mathbf{v}\right)
\end{equation}
where $\gamma=c_p/c_v$ is the ratio of specific heats, generally taken as $1.4$ for air. 

Along wall boundaries we apply the slip boundary condition
\begin{equation}
v_n(w) := \mathbf{v}\cdot {\bf n}=0.
\end{equation}
We also define a boundary function which satisfies $v_n(w_{\partial\Omega}(w))=0$ as
\begin{equation}
w_{\partial\Omega}(w)=\begin{pmatrix}
1 & \mathbf{0}^T & 0 \\
\mathbf{0} & \mathbf{Id}-\mathbf{n}\otimes\mathbf{n}& \mathbf{0} \\
0 & \mathbf{0}^T & 1
\end{pmatrix}w.
\end{equation}

Prescribing boundary conditions at the far-field can be realized with the aid of characteristic upwinding \cite{godlewski1996}. Here, the normal convective flux Jacobian is decomposed as
\begin{equation}
\mathbf{f}_c^{\prime}(w)\cdot\mathbf{n}=Q(w,\mathbf{n})\cdot \Lambda(w,\mathbf{n})\cdot Q^{-1}(w,\mathbf{n})
\end{equation}
with $\Lambda(w,\mathbf{n})$ being a diagonal matrix containing the eigenvalues of $\mathbf{f}_c^{\prime}(w)\cdot\mathbf{n}$. The corresponding right eigenvectors can be found in the columns of $Q(w,\mathbf{n})$. The interior and far-field states in characteristic variables are then given by $w_c=Q(w,\mathbf{n})w$ and $w_{c,\infty}=Q(w,\mathbf{n})w_{\infty}$, respectively. Finally, depending on the sign of $\left(\Lambda(w,\mathbf{n})\right)_{i,i}$, we can construct a boundary state
\begin{equation}
\left(w_{\partial\Omega}(w)\right)_i=\begin{cases}
\left(Q(w,\mathbf{n})w_c\right)_i, & \left(\Lambda(w,\mathbf{n})\right)_{i,i}\geq 0\\
\left(Q(w,\mathbf{n})w_{c,\infty}\right)_i, & \left(\Lambda(w,\mathbf{n})\right)_{i,i}<0.
\end{cases}
\end{equation}

\subsection{Two-Dimensional Navier-Stokes Equations}
The Navier-Stokes equations in conservative form  are given by
\begin{equation}
\boldsymbol\nabla\cdot\left(\mathbf{f}_c(w)-\mathbf{f}_v(w,\boldsymbol\nabla w)\right)=0.
\end{equation}
The convective part $\mathbf{f}_c$ of the Navier-Stokes equations coincides with the Euler equations. The viscous flux is given by 
\begin{align}
\mathbf{f}_v=\left({\bf 0},\boldsymbol\tau, \boldsymbol\tau\mathbf{v}+k\boldsymbol\nabla T\right)^T.
\end{align}

The temperature is defined via the ideal gas law
\begin{equation}
T=\frac{\mu\gamma}{k\cdot\mathrm{Pr}}\left(\frac{E}{\rho}-\frac12\mathbf{v}\cdot\mathbf{v}\right)=\frac{1}{(\gamma-1)c_v}\frac{p}{\rho}
\end{equation}
where $\mathrm{Pr}=\frac{\mu c_p}{k}$ is the Prandtl number, which for air at moderate conditions can be taken as a constant with a value of $\mathrm{Pr}\approx 0.72$. $k$ denotes the thermal conductivity coefficient.
For a Newtonian fluid, the stress tensor is defined as 
\begin{equation}
\boldsymbol\tau=\mu\left(\boldsymbol\nabla\mathbf{v}+\left(\boldsymbol\nabla\mathbf{v}\right)^T-\frac23\left(\boldsymbol\nabla\cdot\mathbf{v}\right)\mathbf{Id}\right).
\end{equation}

The variation of the molecular viscosity $\mu$ as a function of temperature is determined by Sutherland's law as
\begin{equation}
\mu=\frac{C_1T^{3/2}}{T+C_2}
\end{equation}
with $C_1=\SI{1.458e-6}{\frac{kg}{ms\sqrt{K}}}$ and $C_2=\SI{110.4}{K}$.

Along wall boundaries, we apply the no-slip boundary condition, i.e.
\begin{equation}
\mathbf{v}=\mathbf{0}
\end{equation}
with corresponding boundary function
\begin{equation}
w_{\partial\Omega}(w)=\left(\rho, \mathbf{0}, E\right)^T.
\end{equation}
Furthermore, one has to give boundary conditions for the temperature. In the present work we use the adiabatic wall condition, i.e.
\begin{equation}
\boldsymbol\nabla T\cdot\mathbf{n}=0.
\end{equation}
Combining both no-slip and adiabatic wall boundary conditions, yields a condition for the viscous flux, namely
\begin{equation}
\mathbf{f}_v\left(w_{\partial\Omega}, \mathbf{q}_{\partial\Omega}\right)=\begin{pmatrix}
\mathbf{0} & \boldsymbol\tau & \mathbf{0}
\end{pmatrix}^T.
\end{equation}

%% file: discretization.tex

\section{Discretization}
\label{sec:discretization}

\subsection{Notation}

We tesselate the domain $\Omega$ into a collection of non-overlapping elements, denoted by $\mathcal{T}_h$, such that $\bigcup_{K\in\mathcal{T}_h}K=\Omega$. For the element edges we consider two different kinds of sets, $\partial\mathcal{T}_h$ and $\Gamma_h$, which are element-oriented and edge-oriented, respectively. 
{\footnotesize
\begin{alignat}{3}
 \partial \mathcal{T}_h &:= \{ \ \partial K \backslash \partial \Omega \ :\  K \in \mathcal{T}_h \ \}, &\\
 \Gamma_h     &:= \{ \ e \ : \ e = K \cap K'  \text{ for } K, K' \in \mathcal{T}_h; \mathrm{meas}_{d-1}(e)\neq 0 \ \}.
\end{alignat}}
The first is the collection of all element boundaries, which means that every edge appears twice. The latter, however, includes every edge just once. The reason for this distinction will become clear later. Please note that neither of these sets shall include edges lying on the domain boundary; the set of boundary edges is denoted by $\Gamma^b_h$.

We denote by $\Pi^{p}(D)$ the set of polynomials of degree at most $p$ on some domain $D$. We will need discontinuous function spaces for the domain and the mesh skeleton:
{\footnotesize
\begin{alignat}{5}
\mathbf{V}_h &= \{v\in L^2\left(\Omega\right)     &\ :\ & v|_K     &\ \in\ & \Pi^{p_K}(K), &\quad& K&\ \in\ &\mathcal{T}_h\}^{m\times d}\\
W_h &= \{w\in L^2\left(\Omega\right)     &\ :\ & w|_K     &\ \in\ & \Pi^{p_K}(K), &\quad&K&\ \in\ &\mathcal{T}_h\}^m\\
M_h &= \{\mu\in L^2\left(\Gamma_h\right) &\ :\ & \mu|_{e} &\ \in\ & \Pi^{p_e}(e), &\quad&e&\ \in\ &\Gamma_h\}^m.
\end{alignat}}
Thus, $\mathbf{v}\in \mathbf{V}_h$, $w\in W_h$ and $\mu\in M_h$ are piecewise polynomials of degree $p$ which can be discontinuous across edges (for $\mathbf{v}$, $w$) or vertices (for $\mu$), respectively. 

Usually, the polynomial degree between elements and interfaces does not vary. In the case of varying polynomial degrees $p_{K^-}$ and $p_{K^+}$, we follow \citet{cockburn2009unified} and choose the polynomial degree for the interface $e=K^-\cap K^+$ as $p_e=\max\left\{p_{K^-},p_{K^+}\right\}$. This approach is commonly applied for $p$-adaptive HDG discretizations (see \citet{chen2012analysis}, \citet{cockburn2005error} and  \citet{giorgiani2013hybridizable}).

We will distinguish between element-oriented inner products (defined with $\mathcal{T}_h$), i.e.
\begin{align}
\vint{v,w}&:=\sum_{K\in\mathcal{T}_h}\int_K vw\dx, \\
\vint{\mathbf{v},\mathbf{w}}&:=\sum_{K\in\mathcal{T}_h}\int_K \mathbf{v}\cdot\mathbf{w}\dx, \\ 
\eint{v,w}&:=\sum_{K\in\mathcal{T}_h}\int_{\partial K}vw\ds,
\end{align}
and edge-oriented inner products (defined with $\Gamma_h$), i.e.
\begin{equation}
\eints{v,w}:=\sum_{e\in\Gamma_h}\int_e\ vw\ds.
\end{equation}

\subsection{Weak Formulation}
We can rewrite general convection-diffusion equations as a first-order system by introducing an additional unknown representing the gradient of the solution
\begin{align}\begin{split}
\label{eq:strmixed}
\mathbf{q}&=\boldsymbol\nabla w \\
\boldsymbol\nabla\cdot\left(\mathbf{f}_c\left(w\right)-\mathbf{f}_v\left(w,\mathbf{q}\right)\right)&=s\left(w,\mathbf{q}\right).
\end{split}\end{align}

By multiplying the strong, mixed form \eqref{eq:strmixed} with appropriate test functions $\left(\boldsymbol\tau_h,\varphi_h\right)\in \mathbf{V}_h\times W_h$ and integrating by parts, we obtain a standard DG discretization in mixed formulation of the problem, i.e.:

Find $(\mathbf{q}_h,w_h)\in\mathbf{V}_h\times W_h$ s.t. $\forall(\boldsymbol\tau_h,\varphi_h)\in\mathbf{V}_h\times W_h$
\begin{align}
\footnotesize
0=&\phantom{-}\mathcal{N}^{\rm DG}_h\left(\mathbf{q}_h,w_h;\boldsymbol\tau_h,\varphi_h\right)\\
:=&\phantom{-}\vint{\boldsymbol\tau_h,\mathbf{q}_h}+\vint{\boldsymbol\nabla\cdot\boldsymbol\tau_h,w_h}-\eint{\boldsymbol\tau_h\cdot\mathbf{n},\widehat{w}}\\
&-\vint{\boldsymbol\nabla\varphi_h,\mathbf{f}_c(w_h)-\mathbf{f}_v(w_h,\mathbf{q}_h)}\\
&+\eint{\varphi_h, \widehat{f}_c-\widehat{f}_v}-\vint{\varphi_h, s(w_h,\mathbf{q}_h)}\\
&+\mathcal{N}^{\rm DG}_{h,\partial\Omega}\left(\mathbf{q}_h,w_h;\boldsymbol\tau_h,\varphi_h\right)+\mathcal{N}^{\rm DG}_{h,\rm sc}\left(\mathbf{q}_h,w_h;\varphi_h\right).
\end{align}

Here the numerical trace $\widehat{w}$ and the numerical fluxes $\widehat{f}_c, \widehat{f}_v$ have to be chosen appropriately to define a stable and consistent method. Furthermore, boundary conditions and shock-capturing are incorporated using appropriate operators, here denoted by $\mathcal{N}^{\rm DG}_{h,\partial\Omega}\left(\mathbf{q}_h,w_h;\boldsymbol\tau_h,\varphi_h\right)$ and $\mathcal{N}^{\rm DG}_{h,\rm sc}\left(\mathbf{q}_h,w_h;\varphi_h\right)$.

In contrast to a DG discretization, where the numerical trace $\widehat{w}$ is defined explicitly in terms of $w_h$ and $\mathbf{q}_h$, it is treated as an additional unknown in an HDG method. This additional unknown is called $\lambda_h$ and has support on the skeleton of the mesh only.
In order to close the system the continuity of the numerical fluxes across edges is required in a weak sense, resulting in an additional equation.

The weak formulation of the hybrid system, comprised of equations for the gradient $\mathbf{q}_h$, the solution $w_h$ itself and its trace on the mesh skeleton $\lambda_h$, is then given by:

Find $(\mathbf{q}_h,w_h,\lambda_h)\in\mathbb{X}_h:=\mathbf{V}_h\times W_h\times M_h$ s.t. $\forall(\boldsymbol\tau_h,\varphi_h,\mu_h)\in \mathbb{X}_h$
\begin{align}
\footnotesize
\label{eq:hdg_weakform}
0=&\phantom{-}\mathcal{N}_h\left(\mathbf{q}_h,w_h,\lambda_h;\boldsymbol\tau_h,\varphi_h,\mu_h\right)\\
\label{eq:hdg_ls1}
:=&\phantom{-}\vint{\boldsymbol\tau_h,\mathbf{q}_h}+\vint{\boldsymbol\nabla\cdot\boldsymbol\tau_h,w_h}-\eint{\boldsymbol\tau_h\cdot\mathbf{n},\lambda_h}\\
\label{eq:hdg_ls2}
&-\vint{\boldsymbol\nabla\varphi_h,\mathbf{f}_c(w_h)-\mathbf{f}_v(w_h,\mathbf{q}_h)}\\
\label{eq:hdg_ls3}
&+\eint{\varphi_h, \widehat{f}_c-\widehat{f}_v}-\vint{\varphi_h, s(w_h,\mathbf{q}_h)}\\
\label{eq:hdg_ls4}
&+\mathcal{N}_{h,\partial\Omega}\left(\mathbf{q}_h,w_h;\boldsymbol\tau_h,\varphi_h\right)+\mathcal{N}_{h,\rm sc}\left(\mathbf{q}_h,w_h;\varphi_h\right)\\
\label{eq:hdg_cont}
&+\eints{\mu_h,\jump{\widehat{f}_c-\widehat{f}_v}}.
\end{align}

Please note the use of $\partial\mathcal{T}_h$ in the weak formulation of the mixed form (see Eq.~\eqref{eq:hdg_ls1}--\eqref{eq:hdg_ls4}) and $\Gamma_h$ in Eq.~\eqref{eq:hdg_cont} defining $\lambda_h$. This perfectly resembles the character of these equations, being element- and edge-oriented, respectively. The terms tested against $\boldsymbol\tau_h$ and $\varphi_h$ are called local solvers, meaning they do not depend on the solution within neighboring elements but only on the trace of the solution which is approximated by $\lambda_h$. The coupling between elements is then introduced by weakly enforcing the normal continuity of the numerical fluxes across interfaces (see Eq.~\eqref{eq:hdg_cont}).

We choose numerical fluxes similar to the Lax-Friedrich flux and to the LDG flux for the convective and diffusive flux, respectively, i.e.
\begin{align}
\widehat{f}_c\left(\lambda_h,w_h\right)&=\mathbf{f}_c\left(\lambda_h\right)\cdot\mathbf{n}-\alpha_c\left(\lambda_h-w_h\right) \\
\widehat{f}_v\left(\lambda_h,w_h, \mathbf{q}_h\right)&=\mathbf{f}_v\left(\lambda_h, \mathbf{q}_h\right)\cdot\mathbf{n}+\alpha_v\left(\lambda_h-w_h\right)
\end{align}
which can be combined into
\begin{equation}
\widehat{f}_c-\widehat{f}_v=\left(\mathbf{f}_c\left(\lambda_h\right)-\mathbf{f}_v\left(\lambda_h, \mathbf{q}_h\right)\right)\cdot\mathbf{n}-\alpha\left(\lambda_h-w_h\right)
\end{equation}
where $\alpha=\alpha_c+\alpha_v$. The stabilization introduced can be given by a tensor; in our work, however, we restrict ourselves to a constant scalar $\alpha$ which seems to be sufficient for a wide range of test cases.

\subsubsection{Boundary Conditions}
In order to retrieve an adjoint-consistent scheme, special care has to be taken when discretizing the boundary conditions (see \citet{schutz2012adjoint}). The boundary conditions have to be incorporated by using the boundary states $w_{\partial\Omega}\left(w_h\right)$ and gradients $\mathbf{q}_{\partial\Omega}\left(w_h,\mathbf{q}_h\right)$, i.e.
\begin{align}
\footnotesize
&\mathcal{N}_{h,\partial\Omega}\left(\mathbf{q}_h,w_h;\boldsymbol\tau_h,\varphi_h\right)\\
:=&\beint{\boldsymbol\tau_h\cdot\mathbf{n},w_{\partial\Omega}}
+\beint{\varphi_h,\left(\mathbf{f}_c\left(w_{\partial\Omega}\right)-\mathbf{f}_v\left(w_{\partial\Omega},\mathbf{q}_{\partial\Omega}\right)\right)\cdot\mathbf{n}}\nonumber.
\end{align}
We would like to emphasize that $\lambda_h$ does not occur in this boundary term.

\subsubsection{Shock-Capturing}
In non-smooth parts of the solution, for example shocks in compressible flows, a stabilization term has to be introduced. We adopt the shock-capturing approach by \citet{hartmann2002adaptive} where an artificial viscosity term, given by $\boldsymbol\nabla \cdot \left(\epsilon\left(w,\boldsymbol\nabla w\right)\boldsymbol\nabla  w \right)$, is used. The viscosity $\epsilon$ is given by the $L^1$-norm of the strong residual $\boldsymbol\nabla\cdot\mathbf{f}_c(w)$ in every element. In order to accelerate the convergence of this term to zero with mesh refinement, it is premultiplied with an effective mesh size $\tilde{h}_K:=\frac{h_K}{p_K}$. The latter resembles the actual resolution within an element. Furthermore, a user-defined factor $\epsilon_0$ is introduced which can be reliably tuned for a rather large range of test cases. Finally, the artificial viscosity is given by
\begin{equation}
\epsilon|_K:=\frac{\epsilon_0 \tilde{h}_K^{2-\beta}}{|K|}\int_K d(w)\dx
\end{equation}
where the strong residual is given by
\begin{equation}
d(w):=\sum_{i=1}^m\left|(\boldsymbol\nabla\cdot\mathbf{f}_c(w))_i\right|.
\end{equation}

In the discretization of this shock-capturing term the interface integral is neglected so that only the volume contribution is considered, i.e. 
\begin{equation}
\mathcal{N}_{h,\rm sc}\left(w_h;\varphi_h\right):=\vint{\boldsymbol\nabla\varphi_h,\epsilon\left(w_h,\boldsymbol\nabla w_h\right)\boldsymbol\nabla w_h}.
\end{equation}
This obviates the need for introducing $\mathbf{q}_h$ in purely convective problems (e.g. the compressible Euler equations). In the viscous case, where the gradient is explicitly given, $\boldsymbol\nabla w_h$ can be replaced by $\mathbf{q}_h$ yielding
\begin{equation}
\mathcal{N}_{h,\rm sc}\left(\mathbf{q}_h,w_h;\varphi_h\right):=\vint{\boldsymbol\nabla\varphi_h,\epsilon\left(w_h,\mathbf{q}_h\right)\mathbf{q}_h}.
\end{equation}
Please note, that this term enters only the local part of the discretization.

Furthermore, it is noteworthy that this shock-capturing term is not only consistent and conservative but also yields an asymptotically adjoint-consistent scheme.

\subsection{Relaxation}
In order to solve the nonlinear system of equations that defines the HDG method, the Newton-Raphson method is applied. 
Beginning with an initial guess $\mathbbm{x}^0_h := \left({\bf q}_h^0,w_h^0,\lambda_h^0\right)$, one iteratively solves the linear system 
\begin{equation}
\mathcal{N}^{\prime}_h\left[\mathbbm{x}^n_h\right]\left({\delta\mathbbm{x}}^n_h;\mathbbm{y}_h\right)=-\mathcal{N}_h\left(\mathbbm{x}^n_h;\mathbbm{y}_h\right) \quad \forall \mathbbm{y}_h \in \mathbb{X}_h
\label{eq:linearsystem}
\end{equation}
and updates the solution as 
\begin{equation}
\mathbbm{x}^{n+1}_h=\mathbbm{x}^n_h+{\delta\mathbbm{x}}^n_h
\end{equation}
until the residual $\mathcal{N}_h\left(\mathbbm{x}^n_h;\mathbbm{y}_h\right)$ has reached a certain threshold. 
Please note, that we have grouped $\mathbf{q}_h$, $w_h$ and $\lambda_h$, and the test functions into $\mathbbm{x}_h:=\left(\mathbf{q}_h,w_h,\lambda_h\right)$ and $\mathbbm{y}_h:=\left(\boldsymbol\tau_h,\varphi_h,\mu_h\right)$, respectively. 
$\mathcal{N}^{\prime}_h$ denotes the Fr\'echet derivative of $\mathcal{N}_h$ with respect to $\mathbbm{x}_h$.

This routine can, however, lead to stability problems if the starting value $\mathbbm{x}^0_h$ is too far away from the solution $\mathbbm{x}_h$. Therefore, an artificial time is introduced and a backward Euler method is applied, which yields a slight modification of the linear system given in Eq.~\eqref{eq:linearsystem}, namely $\forall \mathbbm{y}_h \in \mathbb{X}_h$
\begin{equation}
\small
\vint{\varphi_h,\frac1{\Delta t^n}\delta{w}^n_h}+\mathcal{N}^{\prime}_h\left[\mathbbm{x}^n_h\right]\left({\delta\mathbbm{x}}^n_h;\mathbbm{y}_h\right)=-\mathcal{N}_h\left(\mathbbm{x}^n_h;\mathbbm{y}_h\right).
\end{equation}
Please note that by choosing $\Delta t^n\rightarrow\infty$, a pure Newton-Raphson method is obtained. Usually the time step is kept finite for a few initial steps to ensure stability. As soon as the residual is lower than a certain threshold, i.e. the current approximation $\mathbbm{x}^n_h$ is thought to be sufficiently close to the solution $\mathbbm{x}_h$, we let the time step go towards infinity.

As we are interested in steady-state problems, we can use local time-stepping in order to accelerate the computation. For each element, we apply a time step based on a global CFL number, the element volume, and an approximation of the flux Jacobian's spectral radius, i.e.
\begin{equation}
\Delta t^n_K={\rm CFL}^n\frac{\left|K\right|}{\lambda_c+4\lambda_v}.
\end{equation}
Here, $\lambda_c$ and $\lambda_v$ represent approximations to the maximum eigenvalue of the convective and diffusive flux, respectively (see \citet{mavriplis1990multigrid}).

The next question is, how to choose $\mathrm{CFL}^n$ in each iteration such that for $\mathbbm{x}^n_h\rightarrow\mathbbm{x}_h$, $\mathrm{CFL}^n\rightarrow\infty$. Therefore, we choose a sequence $\left(\mathrm{CFL}^n\right)_n$ to consist of a ramping phase followed by a very fast increase, i.e.  
\begin{equation}
\label{eq:cfl}
\footnotesize
\mathrm{CFL}^n:=\begin{cases}
c_0\left(3\left(\frac{n}{n_0}\right)^2-2\left(\frac{n}{n_0}\right)^3\right) & n\le n_0 \\
\mathrm{CFL}^{n-1}\left(1+c_1\max{\left(0, \log{\left(\frac{\|\mathcal{N}^{n-1}_h\|_2}{\|\mathcal{N}^n_h\|_2}\right)}\right)}\right) & \text{otherwise}
\end{cases}
\end{equation}
Strictly speaking, the parameters $c_0$, $c_1$ and $n_0$ depend on the actual problem; experiments, however, showed that they can be reliably tuned for a wide range of problems.

\subsection{Hybridization}
Using an appropriate polynomial expansion for $\delta{\mathbf{q}_h}$, $\delta{w_h}$ and $\delta{\lambda_h}$, the linearized global system Eq.~\eqref{eq:linearsystem}
is given in matrix form as
\begin{equation}
\left[\begin{array}{ccc}A & B & R \\C & D & S \\ L & M & N\end{array}\right]\left[\begin{array}{c}\delta Q \\\delta W \\ \delta\Lambda\end{array}\right]=\left[\begin{array}{c}F \\ G \\ H\end{array}\right]
\end{equation}
where the vector $\left[\delta Q, \delta W, \delta\Lambda\right]^T$ contains the expansion coefficients of $\delta{\bf x}_h$ with respect to the chosen basis. 

In order to carry on with the derivation of the hybridized method, we want to formulate that system in terms of $\delta\Lambda$ only. Therefore we split it into
\begin{equation}
\label{eq:ls}
\left[\begin{array}{cc}A & B \\C & D\end{array}\right]\left[\begin{array}{c}\delta Q \\\delta W\end{array}\right]=\left[\begin{array}{c}F \\ G\end{array}\right]-\left[\begin{array}{c}R \\ S\end{array}\right]\delta\Lambda
\end{equation}
and
\begin{equation}
\label{eq:cons}
\left[\begin{array}{cc}L & M\end{array}\right]\left[\begin{array}{c}\delta Q \\\delta W\end{array}\right]+N\delta\Lambda=H.
\end{equation}

Substituting Eq.\,\eqref{eq:ls} into Eq.\,\eqref{eq:cons} yields the hybridized system
\begin{equation}
\label{eq:hybridsystem}
K\,\delta\Lambda=E.
\end{equation}
where the system matrix and right-hand side vector are given by
\begin{align}
K&=\left(N-\left[\begin{array}{cc}L & M\end{array}\right]\left[\begin{array}{cc}A & B \\C & D\end{array}\right]^{-1}\left[\begin{array}{c}R \\S\end{array}\right]\right)\\
E&=H-\left[\begin{array}{cc}L & M\end{array}\right]\left[\begin{array}{cc}A & B \\C & D\end{array}\right]^{-1}\left[\begin{array}{c}F \\ G\end{array}\right].
\end{align}

The workflow is as follows: First, the hybridized system is assembled and then being solved for $\delta\Lambda$. Subsequently, $\delta Q$ and $\delta W$ can be reconstructed inside the elements via Eq.\,\eqref{eq:ls}. It is very important to note that it is not necessary to solve the large system given by Eq.\,\eqref{eq:ls}. 
In fact, the matrix in Eq.\,\eqref{eq:ls} can be reordered to be block diagonal. Each of these blocks is associated to one element. 
Thus, both the assembly of the hybridized matrix in Eq.\,\eqref{eq:hybridsystem} and the reconstruction of $\delta Q$ and $\delta W$ can be done in an element-wise fashion. In order to save computational time, the solutions to Eq.\,\eqref{eq:ls} can be saved after the assembly of the hybridized system and reused during the reconstruction of $\delta Q$ and $\delta W$.

The hybridized matrix is an $n_f\times n_f$ block matrix, where $n_f=|\Gamma_h|$ is the number of interior edges. In each block row there is one block on the diagonal and $2d$ off-diagonal blocks in the case of simplex elements. These blocks represent the edges of the neighboring elements of one edge. Each block is dense and has $\mathcal{O}\left(m^2\cdot p^{2(d-1)}\right)$ entries. Please recall that $p$ is the polynomial degree of the ansatz functions, $d$ is the spatial dimension of the domain $\Omega$ and $m$ is the number of partial differential equations ($m=4$ for the 2-dimensional Euler or Navier-Stokes equations). This structure is very similar to that of a normal DG discretization, whereas the blocks in the latter  have $\mathcal{O}\left(m^2\cdot p^{2d}\right)$ entries and thus are considerably bigger for higher polynomial order $p$. The size of the system matrix does not only play a big role in terms of memory consumption but also for the iterative solver. Here, a major portion of the overall 
workload goes into matrix-vector products which can obviously be performed faster, if the problem dimensions are smaller. In our code we use an ILU($n$)-preconditioned GMRES which is available through the PETSc library \cite{petsc-web-page, petsc-user-ref}.

%% file: errorestimation.tex

\section{Adaptation Procedure}
\label{sec:errorestimation}
In the context of adjoint-based (also referred to as target- or output-based) error estimation, one is interested in quantifying the error of a specific target functional $J_h:\mathbb{X}_h\rightarrow\mathbb{R}$, i.e.
\begin{equation}
e_h:=J_h\left(\mathbbm{x}\right)-J_h\left(\mathbbm{x}_h\right),
\end{equation}
where $\mathbbm{x}_h$ is the approximation to $\mathbbm{x}$ in $\mathbb{X}_h$. This target functional can for example represent lift or drag coefficients in aerospace applications. 
In general, the target functional is an integrated value, where integration can be both on a volume or along the boundary. 
For the derivation of the adjoint-based error estimate we expand the target functional in a Taylor series as follows
\begin{equation}
J_h\left(\mathbbm{x}\right)-J_h\left(\mathbbm{x}_h\right)=J_h^{\prime}\left[\mathbbm{x}_h\right]\delta\mathbbm{x}_h+\mathcal{O}\left(\|\delta\mathbbm{x}_h\|^2\right)
\label{eq:linerr}
\end{equation}
where we defined $\delta\mathbbm{x}_h:=\mathbbm{x}-\mathbbm{x}_h$.

We proceed in a similar manner with the error in the residual, i.e.
\begin{equation}
\footnotesize
\mathcal{N}_h\left(\mathbbm{x}; \mathbbm{y}_h\right)-\mathcal{N}_h\left(\mathbbm{x}_h;\mathbbm{y}_h\right)=\mathcal{N}_h^{\prime}\left[\mathbbm{x}_h\right]\left(\delta\mathbbm{x}_h;\mathbbm{y}_h\right)+\mathcal{O}\left(\|\delta\mathbbm{x}_h\|^2\right).
\label{eq:linres}
\end{equation}
As our discretization is consistent the first term $\mathcal{N}_h\left(\mathbbm{x}; \mathbbm{y}_h\right)$ vanishes.

Substituting Eq.~\eqref{eq:linres} into Eq.~\eqref{eq:linerr} and neglecting the quadratic terms yields 
\begin{equation}
e_h\approx\eta:=-\mathcal{N}_h\left(\mathbbm{x}_h;\mathbbm{z}_h\right)
\label{eq:errest}
\end{equation}
where $\mathbbm{z}_h$ is defined by the so-called adjoint equation
\begin{equation}
\mathcal{N}^{\prime}_h\left[\mathbbm{x}_h\right]\left(\mathbbm{y}_h;\mathbbm{z}_h\right)=J_h^{\prime}\left[\mathbbm{x}_h\right]\left(\mathbbm{y}_h\right)\qquad\forall\mathbbm{y}_h\in\widetilde{\mathbb{X}}_h.
\label{eq:adjoint}
\end{equation}

The adjoint solution $\mathbbm{z}_h=\left(\widetilde{q}_h, \widetilde{w}_h, \widetilde{\lambda}_h\right)\in\widetilde{\mathbb{X}}_h$ represents the link between variations in the residual and in the target functional.

The global error estimate $\eta$ can then be restricted to a single element to yield a local indicator to drive an adaptation procedure, i.e.
\begin{equation}
\eta_K:=\left|\left.\mathcal{N}_h\left(\mathbf{q}_h,w_h,\lambda_h;\widetilde{\mathbf{q}}_h,\widetilde{w}_h,0\right)\vphantom{\big|}\right|_K\right|.
\end{equation}
Here, we want to emphasize that, in contrast to the global error estimate, we ignore the contribution of the hybrid adjoint variable $\widetilde{\lambda}_h$. We found that by taking the whole adjoint into account, jumps across element interfaces were overly penalized. This deserves a more in-depth analysis.

Please note, that the functionals $\mathcal{N}_h$ and $J_h$ and their jacobians have to be evaluated in a somewhat richer space than $\mathbb{X}_h$, namely $\widetilde{\mathbb{X}}_h\supset \mathbb{X}_h$. Otherwise the weighted residual $\mathcal{N}_h\left(\mathbbm{x}_h;\mathbbm{z}_h\right)$ would be identical zero as
\begin{equation}
\mathcal{N}_h\left(\mathbbm{x}_h;\mathbbm{y}_h\right)=0\qquad\forall\mathbbm{y}_h\in \mathbb{X}_h.
\end{equation}
This can be achieved by either mesh refinement or a higher polynomial degree of the ansatz functions. In our setting, especially when using a hierarchical basis, the latter is advantageous with respect to implementational effort and efficiency.

\subsection{Hybridization}
In matrix form, the adjoint system (see Eq.~\eqref{eq:adjoint}) reads as follows
\begin{equation}
\left[\begin{array}{ccc}A & B & R \\C & D & S \\ L & M & N\end{array}\right]^T\left[\begin{array}{c}\widetilde{Q} \\\widetilde{W} \\ \widetilde{\Lambda}\end{array}\right]=\left[\begin{array}{c}\widetilde{F} \\ \widetilde{G} \\ \widetilde{H}\end{array}\right]
\end{equation}
Please note, that in our formulation $\widetilde{H}=0$ as $\lambda_h$ is not defined on the boundary and thus the target functional depends only on $w_h$ and $q_h$.

As the overall structure of the adjoint equation is similar to the primal system (see Eq.~\eqref{eq:hybridsystem}), one can also apply static condensation to the adjoint system which then yields its hybridized form:
\begin{equation}
K^T\,\widetilde{\Lambda}=\widetilde{E}
\end{equation}
where the right-hand side is given by
\begin{equation}
\widetilde{E}=-\left[\begin{array}{cc}R^T & S^T\end{array}\right]\left[\begin{array}{cc}A & B \\C & D\end{array}\right]^{-T}\left[\begin{array}{c}\widetilde{F} \\ \widetilde{G}\end{array}\right].
\end{equation}

It is interesting to note that the hybridized adjoint system matrix is also the transpose of the hybridized primal system matrix (for a higher polynomial order). This is very beneficial for the implementation as the routines for the assembly of this matrix are already available.

The adjoint solution within each element can then be computed with the aid of the adjoint local system, given by
\begin{equation}
\left[\begin{array}{cc}A & B \\C & D\end{array}\right]^T\left[\begin{array}{c}\widetilde{Q} \\\widetilde{W}\end{array}\right]=\left[\begin{array}{c}\widetilde{F} \\ \widetilde{G}\end{array}\right]-\left[\begin{array}{cc}L & M\end{array}\right]^T\widetilde{\Lambda}
\end{equation}
where the matrix is also the transpose of the primal local matrix (see Eq.~\eqref{eq:ls}).

\subsection{Marking Elements for Refinement}
After having obtained a localized error estimate, we have to choose a set of elements to be refined. This can be done in many different ways. 
We adopt a strategy proposed by \citet{dorfler1996convergent}. The aim of this marking strategy is to find the smallest set $\mathcal{M}\subseteq\mathcal{T}_h$ such that the error contributed by this set represents a certain fraction of the total error, i.e.
\begin{equation}
\eta_{\mathcal{M}}\geq \left(1-\theta\right)\eta_{\mathcal{T}_h}.
\end{equation}
The user-defined parameter $\theta$ is of course problem dependent. It can, however, be tuned for a large range of test cases. Please note, that we define the error of any subset of $\mathcal{T}_h$ as $\eta^2_{\mathcal{M}}:=\sum_{K\in\mathcal{M}}\eta^2_K$.

\subsection{Choosing between $h$- and $p$-Adaptation}

\begin{figure}
\centering
\subfloat[Coarse mesh]{
\includegraphics[width=\columnwidth]{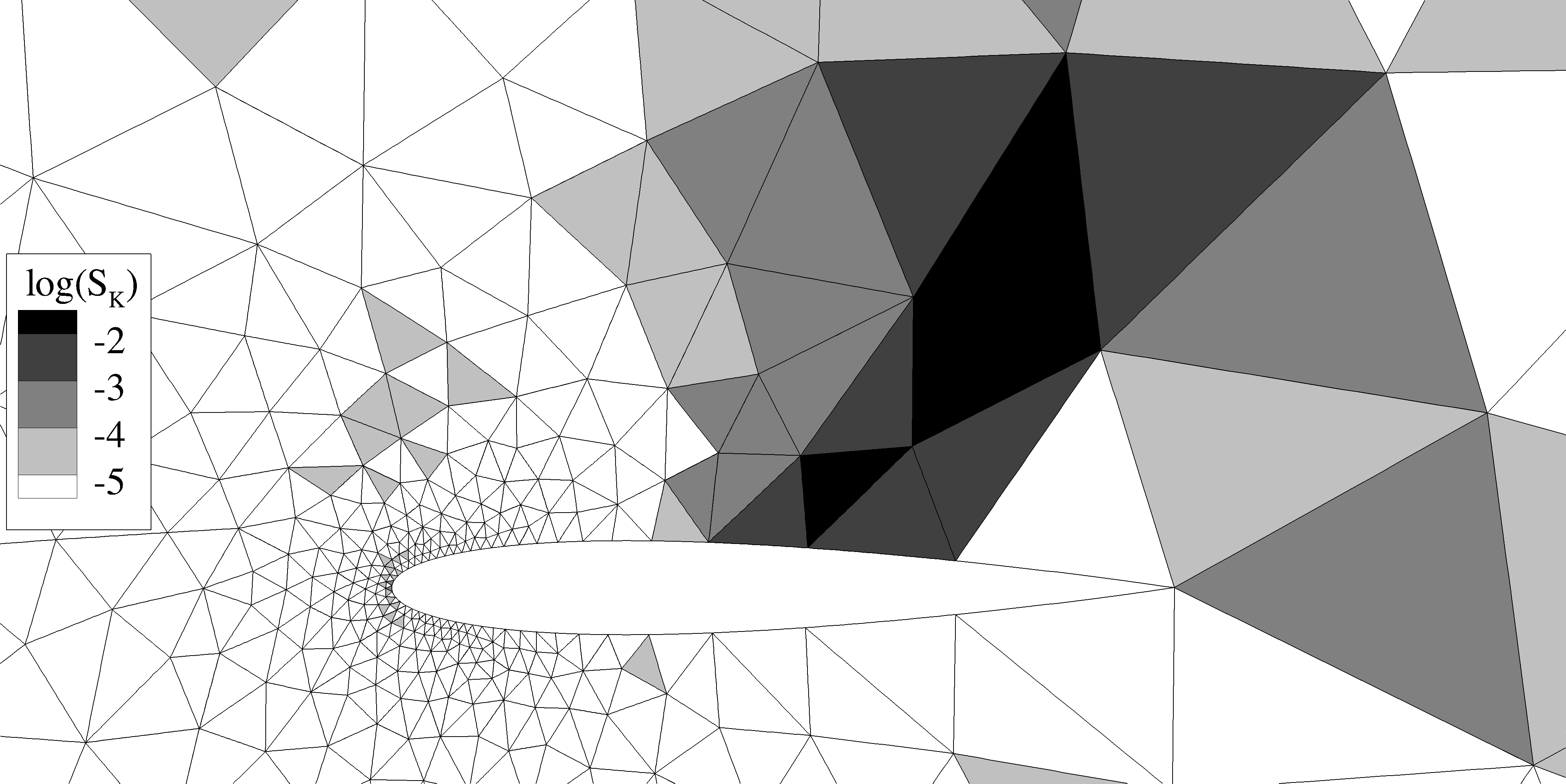}}\\
\subfloat[Adapted mesh]{
\includegraphics[width=\columnwidth]{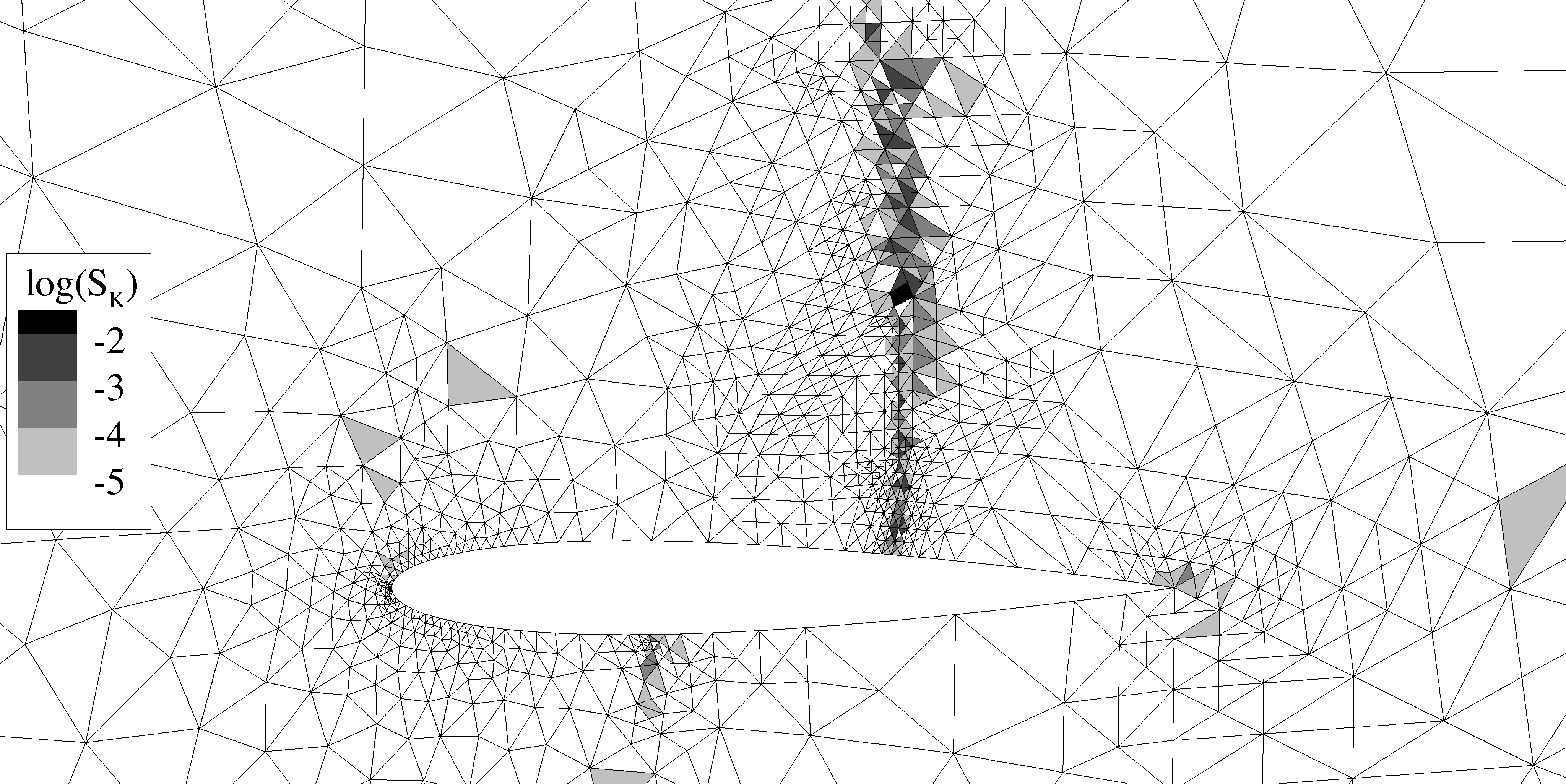}}
\caption{Smoothness sensor for a transonic test case}
\label{fig:smoothness}
\end{figure}

The final step in the adaptation procedure is the decision between mesh refinement and order enrichment. There exist several ways to make this decision. \citet{ceze2010output} solve local adjoint problems for both options and then decide which one is more efficient with respect to degrees of freedom or the non-zero entries in the system matrix. We, however, adopt the strategy by \citet{wangjcp2009}. They used a smoothness sensor originally devised by \citet{persson2006shockcapturing} for an artificial viscosity approach. 
This sensor exploits the fact that the decay of the expansion coefficients of the solution is closely linked to its smoothness. For smoother solutions, decay is faster. 
This can be exploited to check the regularity of the solution. On each element, the smoothness sensor is defined as 
\begin{equation}
S_K :=\frac{\left(w-w^{\star}, w-w^{\star}\right)_K}{\left(w,w\right)_K}
\end{equation}
where $w^{\star}$ is the element-wise projection of $w$ to the next smaller polynomial space, given by
\begin{equation}
\left(\varphi_h,w^{\star}\right)_K=\left(\varphi_h,w\right)_K\qquad \forall \varphi_h\in\Pi^{p_K-1}(K).
\end{equation}
Hence, $w-w^{\star}$ represents the higher order components of the solution (see Fig.~\ref{fig:smoothness}). As we use a hierarchical basis~\cite{dubiner:1991triagTensorBasis}, this projection is very cheap. By introducing a threshold $\epsilon_S$, a decision between mesh-refinement and $p$-enrichment can be made, i.e.
\begin{equation}
S_K\begin{cases}
<\epsilon_S\qquad \text{$p$-enrichment}\\
\geq\epsilon_S\qquad \text{mesh-refinement}
\end{cases}
\end{equation}
First promising results for this approach are given in \cite{balan2013hybridadjointhp}.

%% file: results.tex
\section{Results}
\label{sec:results}

In the following we compare our in-house HDG and DG solvers in terms of degrees
of freedom and runtime. The DG discretization is based on the Lax-Friedrich and
the BR2 \cite{bassi2005discontinuous} fluxes for convective and viscous terms,
respectively. Boundary conditions and target functionals are evaluated in an
adjoint-consistent manner \cite{hartmann2007adjoint}. Both solvers share the
same computational framework, so we believe that our comparison is meaningful. 

We apply both solvers to compressible flow problems, including inviscid subsonic, transonic and supersonic, as well as subsonic laminar flow. In all cases we show results for 
pure mesh-adaptation ($p=1\dots 4$) and $hp$-adaptation ($p=2\dots 5$).

Please note, that we choose the same relaxation parameters for all test cases,
namely, $c_0=10^6$, $c_1=10^3$, $n_0=4$ (see Eq.~\eqref{eq:cfl}). If artificial
viscosity is necessary, we use $\epsilon_0=0.2$ and $\beta=0$. The parameters for
the adaptation procedure are chosen as $\theta=0.05$ and $\epsilon_S=10^{-6}$.
This set of parameters seems to be robust for both DG and HDG for a broad range
of test cases. In order to approximate the error in drag, reference solutions on
$hp$-adapted meshes with more than $2\cdot 10^6$ degrees of freedom are used. (Please
note, that we refer to ${\rm ndof}_w$ whenever we speak of degrees of freedom in
the following as this is a good measure for the resolution.)

\begin{figure}	
\subfloat[Euler]{\includegraphics[width=0.5\columnwidth]{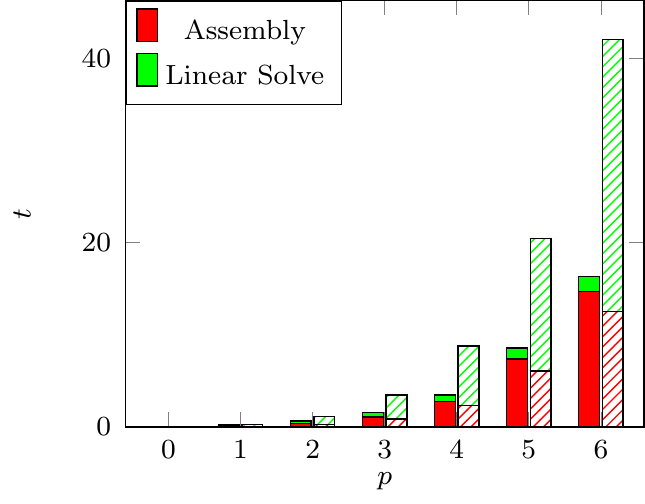}}
\subfloat[Navier-Stokes]{\includegraphics[width=0.5\columnwidth]{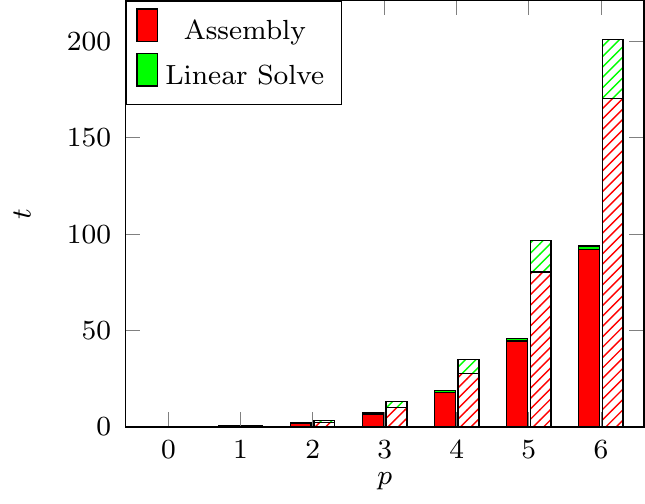}}
\caption{Runtime comparison of the hybridized and non-hybridized DG method for a
         fixed mesh and varying polynomial degree (Solid: HDG, Shaded: DG).}
\label{fig:comparisontime}
\end{figure}

Before we turn our attention to the adaptive computations, we  compare
runtimes for both methods on a fixed mesh for several polynomial orders. This
way, we can learn which improvement can be expected. In
Fig.~\ref{fig:comparisontime} timings for the assembly procedure and the
iterative solver are given. We show Euler and a Navier-Stokes computation on a
mesh with 2396 elements and 3544 interior faces. We used polynomial orders from
$p=0$ to $p=6$. We want to emphasize that the necessary Newton and GMRES
iterations for both HDG and DG were comparable. As there are more
faces than elements, DG is faster than HDG for $p=0$ and $p=1$. However,
already for $p=2$ HDG outruns DG. At $p=6$ there is a ratio of 2.5 for the
Euler test case and 2.1 for the Navier-Stokes test case. For the Euler test
case it is interesting to note, that the iterative solver dominates the
computational time for DG. For HDG it is the other way around. This has two
reasons: firstly, the assembly is more involved due to the local solves; and
secondly, the global system is considerably smaller for higher polynomial
degree. In the case of a Navier-Stokes computation, the DG assembly takes over
the dominating part as the lifting operators are very expensive to compute.
The time for the HDG assembly is also increased which is among other things
due to the introduction of the gradient. For both HDG and DG, the time spent
in the iterative solver is comparable for both Euler and Navier-Stokes.
	
\subsection{Subsonic Inviscid Flow over the NACA~0012 Airfoil}
\begin{figure}
\centering
\includegraphics[width=\columnwidth]{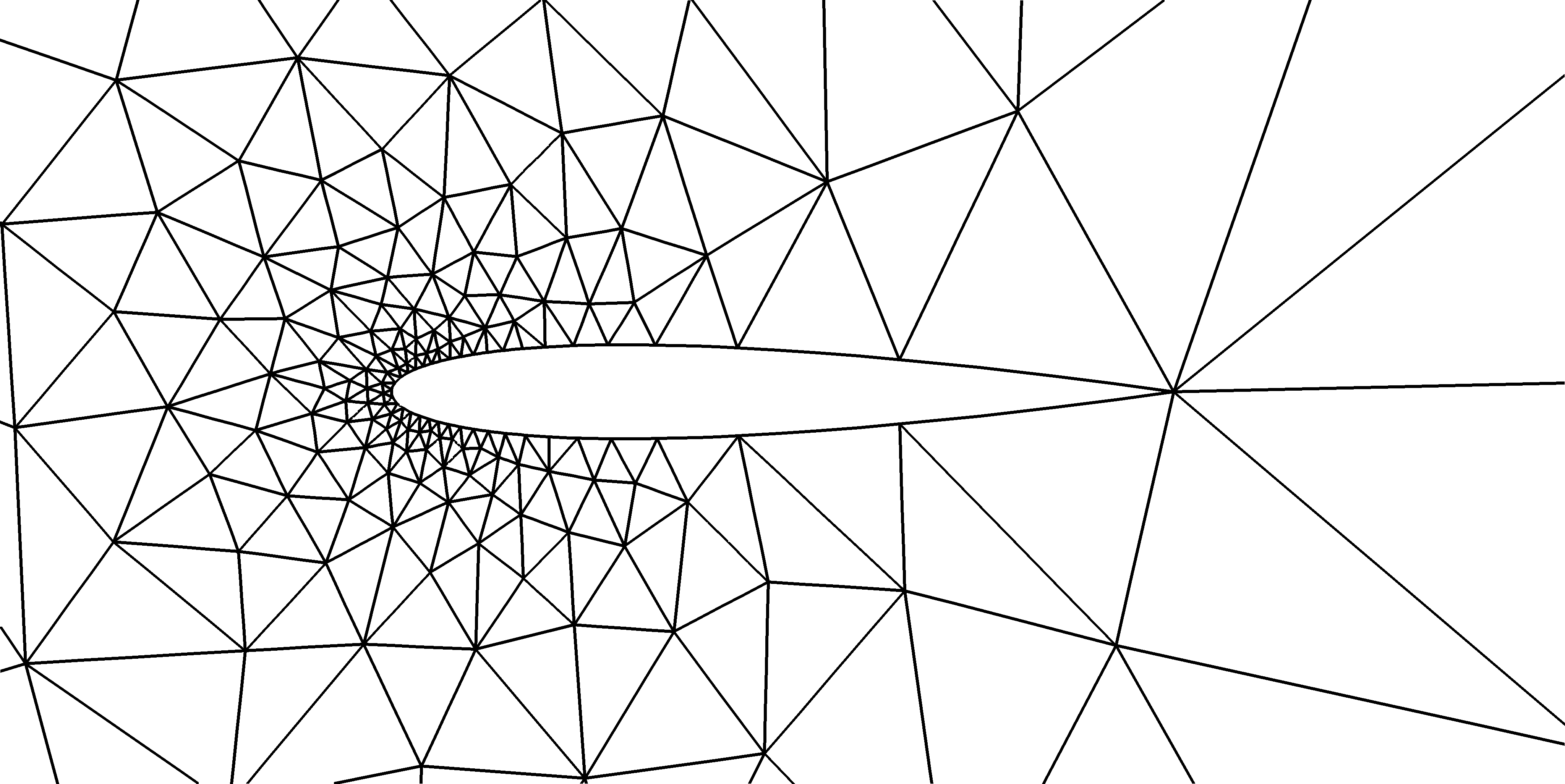}
\caption{Baseline mesh with 719 elements for inviscid computations}
\label{fig:euler_mesh_coarse}
\end{figure}

In the first test case, we consider subsonic inviscid flow over the NACA~0012
airfoil which is defined by
{\footnotesize\begin{equation*}
y=\pm 0.6\left(0.2969\sqrt{x}-0.1260x-0.3516x^2+0.2843x^3-0.1015x^4\right)
\end{equation*}}\noindent with $x\in [0,1]$.
Using this definition, the airfoil would have a finite
trailing edge thickness of .252~\%. In order to obtain a sharp trailing edge we
modify the $x^4$ coefficient, i.e.
{\footnotesize\begin{equation*}
y=\pm 0.6\left(0.2969\sqrt{x}-0.1260x-0.3516x^2+0.2843x^3-0.1036x^4\right).
\end{equation*}}
The flow is characterized by a free stream Mach number of $\rm Ma_{\infty}=0.5$
and an angle of attack of $\alpha=2^{\circ}$. In Fig.~\ref{fig:euler_mesh_coarse}
the baseline mesh for the Euler test cases (subsonic and transonic) can be seen.
It consists of 719 elements and its far field is over a 1000 chords away. 

Admissible target functionals defined on the boundary for the Euler equations are
given by the weighted pressure along wall boundaries, i.e.
\begin{equation}
J(w)=\int_{\partial\Omega}\boldsymbol\psi\cdot\left(p\mathbf{n}\right)\ds
\end{equation}
where $\mathbf{n}$ is the outward pointing normal. By using
$\boldsymbol\psi=\frac1{C_{\infty}}\left(\cos{\alpha}, \sin{\alpha}\right)^T$ or
$\boldsymbol\psi=\frac1{C_{\infty}}\left(-\sin{\alpha}, \cos{\alpha}\right)^T$
along wall boundaries and 0 otherwise, the functional represents the pressure
drag coefficient $c_D$ or the pressure lift coefficient $c_L$, respectively.
$C_{\infty}$ is a normalized reference value defined by
$C_{\infty}=\frac12\gamma{\rm Ma}^2_{\infty}p_{\infty}l$. Here, $l$ is the chord
length of the airfoil.

\begin{figure}
\centering
\subfloat[Pure $h$-adapation ($p=2$)\label{fig:naca_ma05_al2_mesh_h}]{
\includegraphics[width=\columnwidth]{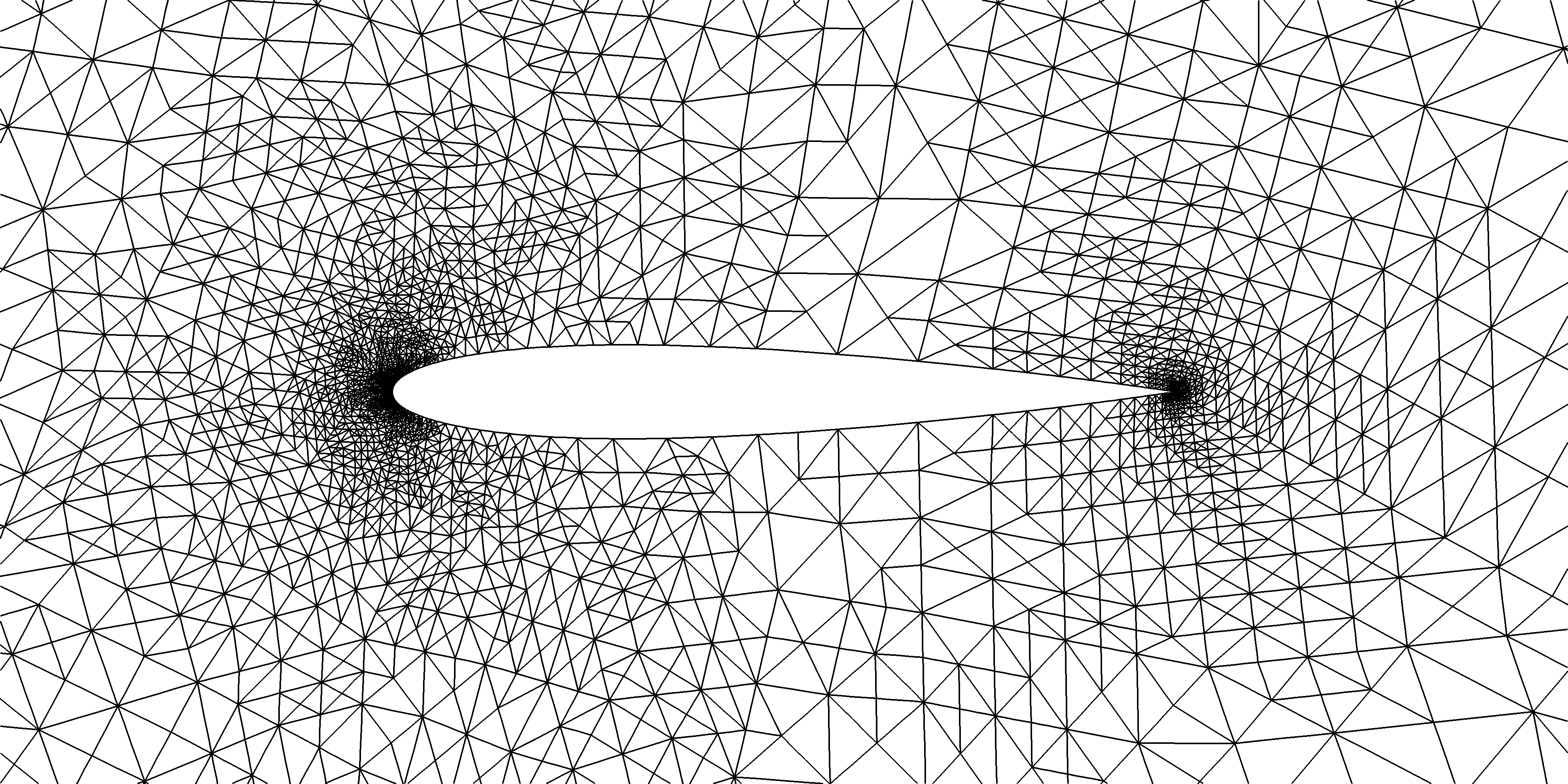}}\\
\subfloat[$hp$-adapation ($p=2\dots 5$)\label{fig:naca_ma05_al2_mesh_hp}]{
\includegraphics[width=\columnwidth]{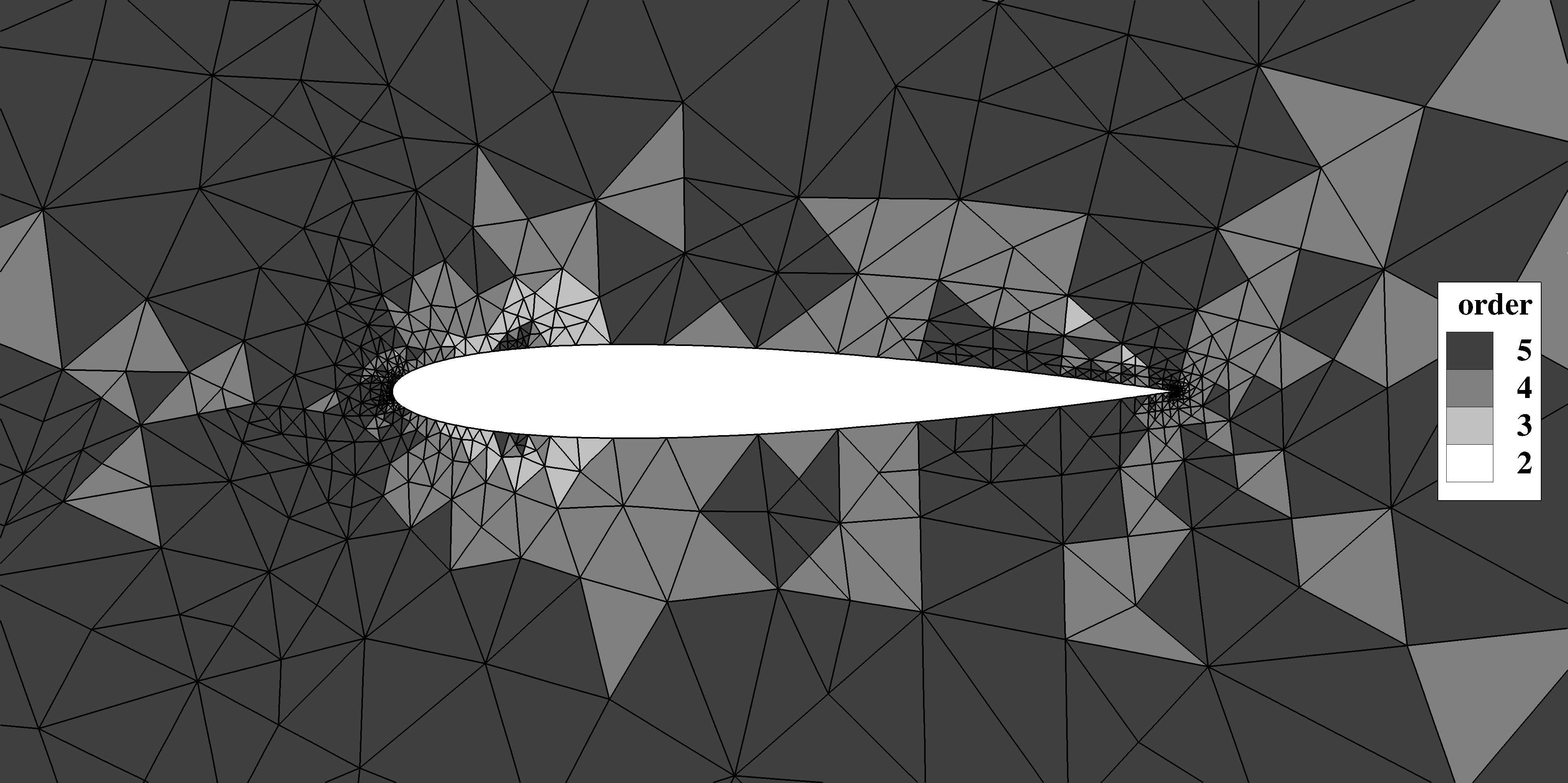}}
\caption{Adapted meshes for the subsonic Euler test case (${\rm Ma}_{\infty}=0.5$, $\alpha=2^{\circ}$)}
\end{figure}

In Fig.~\ref{fig:naca_ma05_al2_mesh_h}, a purely $h$-adapted mesh can be seen.
The most refined regions are the leading and trailing edge. The former is of
importance as the flow experiences high gradients towards the stagnation point.
Refinement of the latter is necessary due to the sharp trailing edge and the
slip-wall boundary conditions. As soon as the error in these two regions is
sufficiently low, other elements close to the airfoil get refined as well. For
the $hp$-adapted mesh (see Fig.~\ref{fig:naca_ma05_al2_mesh_hp}) the leading
and trailing edge are refined as well. All other regions, however, undergo
mostly $p$-enrichment.

\begin{figure}[h]
  \subfloat[HDG]{\includegraphics[width=0.5\columnwidth]{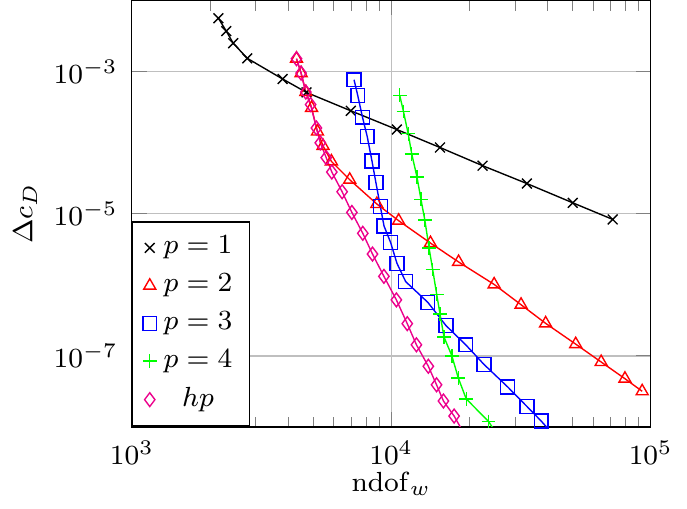}}
  \subfloat[DG]{\includegraphics[width=0.5\columnwidth]{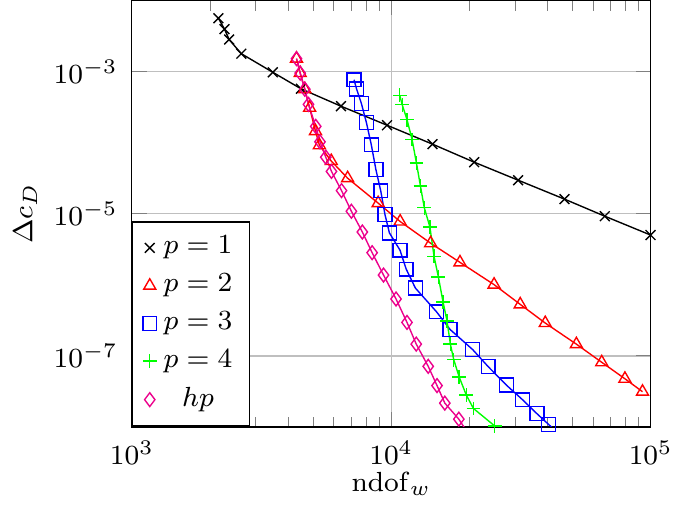}}
  \caption{Drag convergence with respect to degrees of freedom (${\rm Ma}_{\infty}=0.5$, $\alpha=2^{\circ}$)}
  \label{fig:ma05al2_ndof}
\end{figure}

\begin{figure}[h]
  \subfloat[HDG]{\includegraphics[width=0.5\columnwidth]{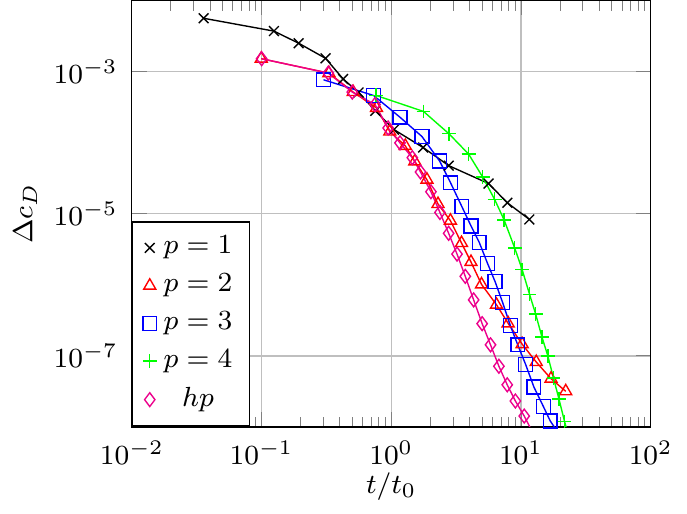}}
  \subfloat[DG]{\includegraphics[width=0.5\columnwidth]{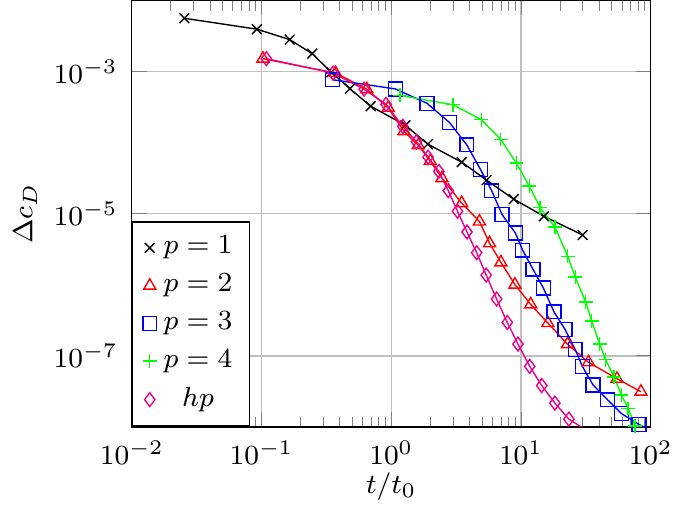}}
  \caption{Drag convergence with respect to time (${\rm Ma}_{\infty}=0.5$, $\alpha=2^{\circ}$)}
  \label{fig:ma05al2_time}
\end{figure}

In terms of degrees of freedom, both HDG and DG show similar results. For all
computations it takes some adaptations until the critical regions, leading and
trailing edge, are resolved. From this point on, one can see the benefit of a
higher order discretization: the error drops significantly faster with respect
to degrees of freedom and computational time (see Fig.~\ref{fig:ma05al2_ndof}
and \ref{fig:ma05al2_time}).

In Tbl.~\ref{tbl:runtime_subsonic}, we give the runtime ratios for a fixed error level (we always choose the minimum level attained). For $p=1$, HDG and DG are comparable in both runtime and nonzero entries. From $p=2$ on, HDG is faster than DG and needs less memories for the global system. Here, it is important to note that the adjoint is approximated with $p+1$, so that already for lower orders HDG is faster. Furthermore, we list the ratio of nonzero entries in the system matrix in order to compare the memory requirements of both methods. As expected, the ratio grows with increasing polynomial degree in favor of HDG.

\begin{table}
\centering
\begin{tabular}{c|c|c|c|c|c}
\hline 
$p$ & $1$ & $2$ & $3$ & $4$ & $hp$\\ \hline
$t_{\rm DG}/t_{\rm HDG}$ & 1.293 & 3.799 & 3.551 & 3.092 & 2.209 \\
$n_{\rm nz, DG}/n_{\rm nz, HDG}$ & 1.123 & 2.150 & 3.248 & 4.281 & 4.601 \\
\hline
\end{tabular}
\caption{Runtime and nonzero ratios for a fixed error level ($\rm Ma_{\infty}=0.5$, $\alpha=2^{\circ}$)}
\label{tbl:runtime_subsonic}
\end{table}

\subsection{Transonic Inviscid Flow over the NACA~0012 Airfoil}

Next, we turn our attention to transonic flow which develops more complex features (e.g.\ compression shocks) compared to the subsonic regime. The flow is characterized by a free stream Mach number of $\rm Ma_{\infty}=0.8$ and an angle of attack of $\alpha=1.25^{\circ}$.

\begin{figure}
\centering
\subfloat[Pure $h$-adaptation ($p=2$)\label{fig:naca_ma08_al125_mesh_h}]{
\includegraphics[width=\columnwidth]{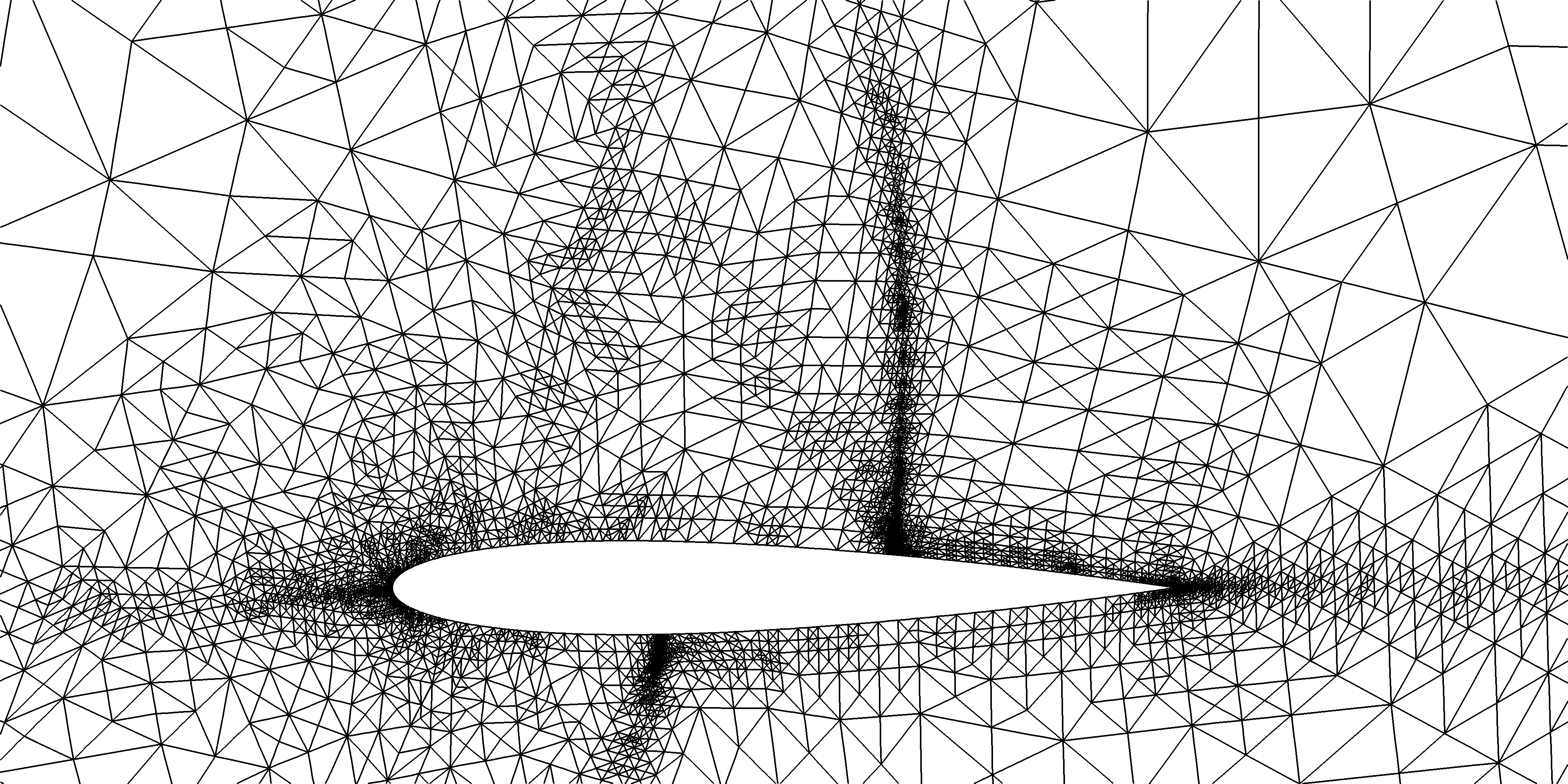}}\\
\subfloat[$hp$-adaptation ($p=2\dots 5$)\label{fig:naca_ma08_al125_mesh_hp}]{
\includegraphics[width=\columnwidth]{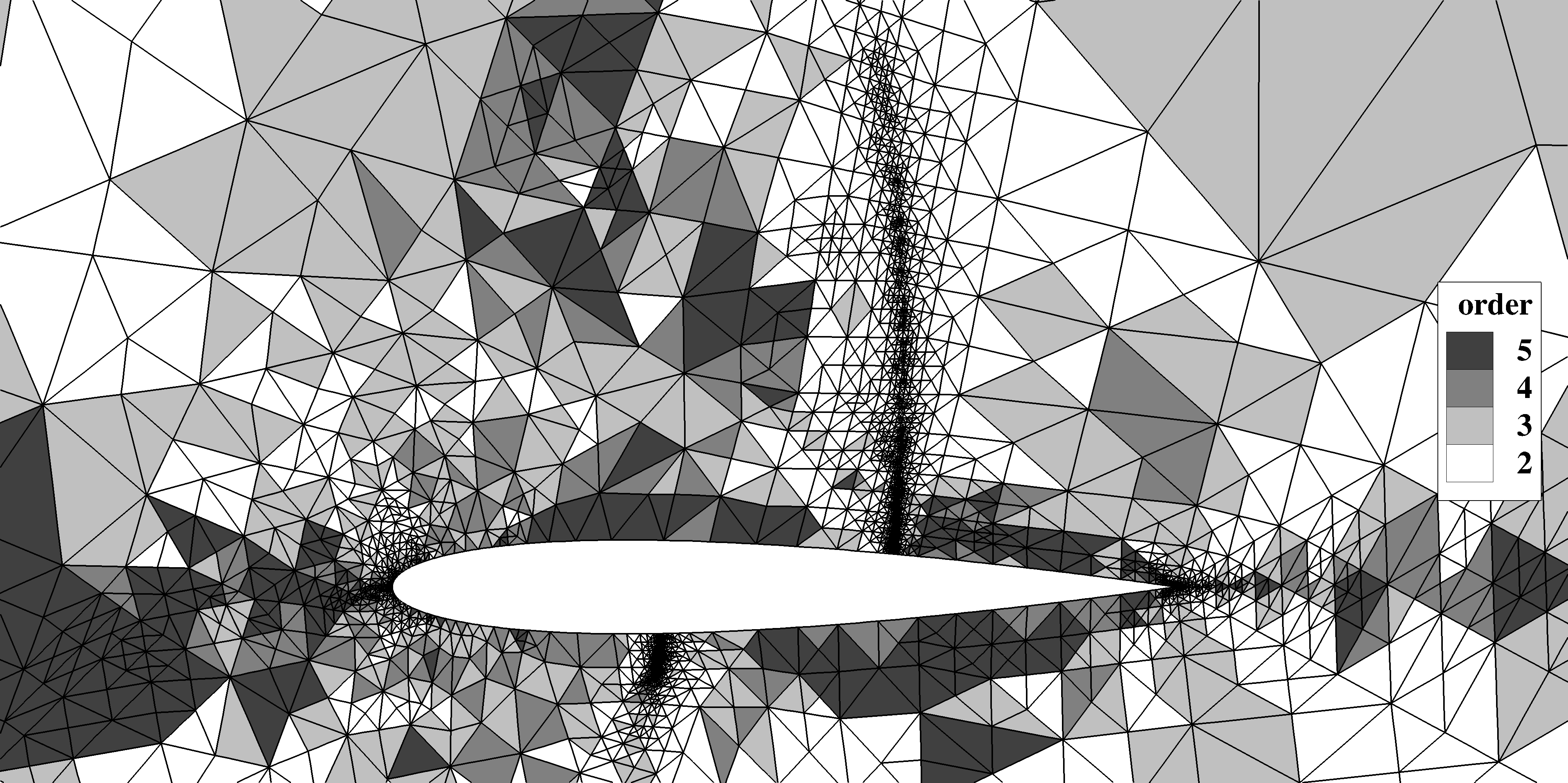}}
\caption{Adapted meshes for the transonic Euler test case (${\rm Ma}_{\infty}=0.8$, $\alpha=1.25^{\circ}$)}
\end{figure}

In Fig.~\ref{fig:naca_ma08_al125_mesh_h} a purely $h$-adapted mesh can be seen. The adjoint sensor detects all regions of relevance for the drag: the upper shock, the leading and trailing edge, and the lower weak shock. Further refinement is added upstream of the shock, where the adjoint has steep gradients and thus needs higher resolution. In the case of $hp$-adaptation, the mesh-refinement is stronger confined to the shock region and the trailing edge. The other features undergo $p$-enrichment.

\begin{figure}[h]
  \subfloat[HDG]{\includegraphics[width=0.5\columnwidth]{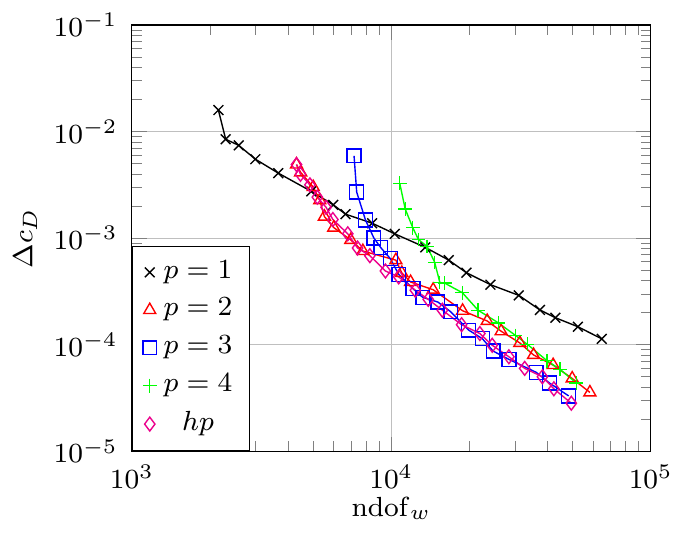}}
  \subfloat[DG]{\includegraphics[width=0.5\columnwidth]{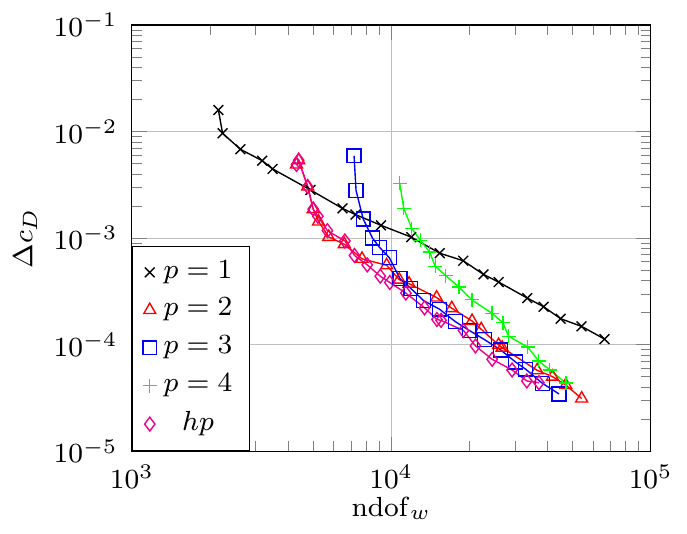}}
  \caption{Drag convergence with respect to degrees of freedom (${\rm Ma}_{\infty}=0.8$, $\alpha=1.25^{\circ}$)}
  \label{fig:ma08al125_ndof}  
\end{figure}

\begin{figure}[h]
  \subfloat[HDG]{\includegraphics[width=0.5\columnwidth]{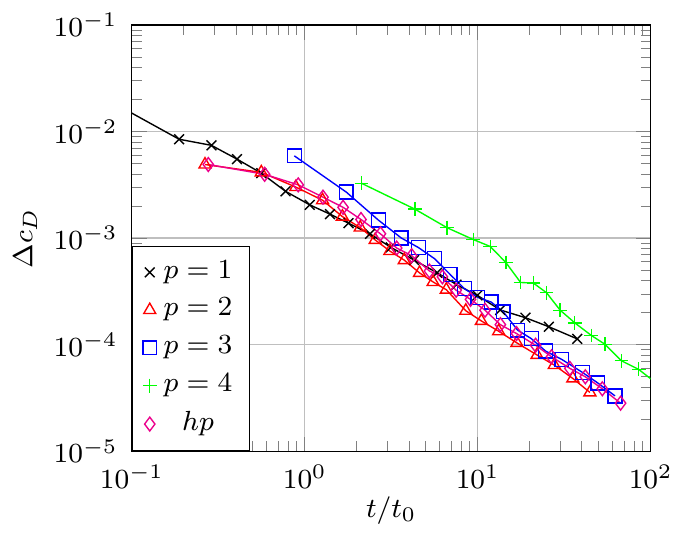}}
  \subfloat[DG]{\includegraphics[width=0.5\columnwidth]{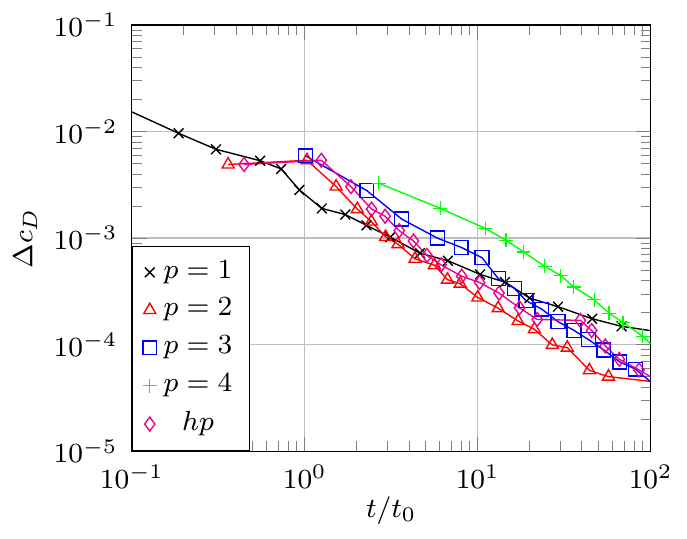}}
  \caption{Drag convergence with respect to time (${\rm Ma}_{\infty}=0.8$, $\alpha=1.25^{\circ}$)}
  \label{fig:ma08al125_time}  
\end{figure}

As expected, both methods show a similar accuracy for a given number of degrees of freedom. The computations with $p=2\dots 4$ outperform $p=1$ but are comparable to each other. $hp$-adaptation shows very good results which is due to the accurate prediction of the solution smoothness (see Fig.~\ref{fig:ma08al125_ndof} and \ref{fig:ma08al125_time}).

For this test case, HDG is faster than DG from $p=1$ on (see Tbl.~\ref{tbl:runtime_transonic}). The $hp$-adaptive run is more than 2.5 times faster. The ratio of necessary nonzero entries attains its highest value for $p=4$. In the case of $hp$-adaptation, this ratio is not as high, as in the shock region a lot of elements with $p=2$ exist.

\begin{table}
\centering
\begin{tabular}{c|c|c|c|c|c}
\hline 
$p$ & $1$ & $2$ & $3$ & $4$ & $hp$\\ \hline
$t_{\rm DG}/t_{\rm HDG}$ & 2.636 & 1.607 & 2.154 & 2.362 & 2.628 \\
$n_{\rm nz, DG}/n_{\rm nz, HDG}$ & 1.237 & 1.807 & 3.098 & 4.437 & 2.411 \\
\hline 
\end{tabular}
\caption{Runtime and nonzero ratios for a fixed error level ($\rm Ma_{\infty}=0.8$, $\alpha=1.25^{\circ}$)}
\label{tbl:runtime_transonic}
\end{table}

\subsection{Supersonic Inviscid Flow over the NACA~0012 Airfoil}

Now, we consider supersonic, inviscid flow with a free stream Mach number of $\mathrm{Ma}=1.4$. We choose the angle of attack to be $\alpha=0^{\circ}$. In contrast to the transonic test case, a so-called bow shock develops in front of the airfoil. 

\begin{figure}
\centering
\subfloat[Pure $h$-adapation ($p=2$)\label{fig:naca_ma14_al0_mesh_h}]{
\includegraphics[width=\columnwidth]{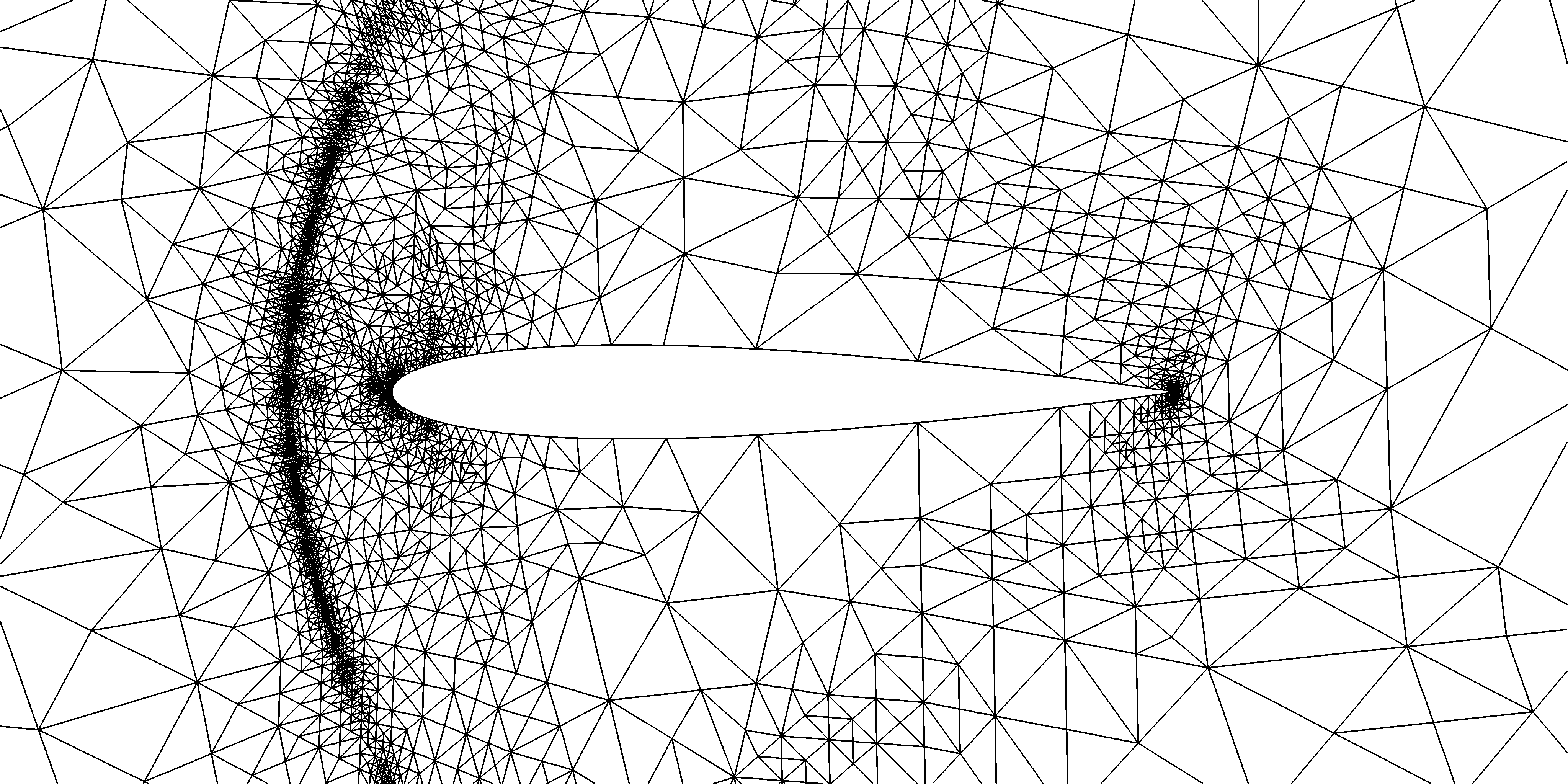}}\\
\subfloat[$hp$-adapation ($p=2\dots 5$)\label{fig:naca_ma14_al0_mesh_hp}]{
\includegraphics[width=\columnwidth]{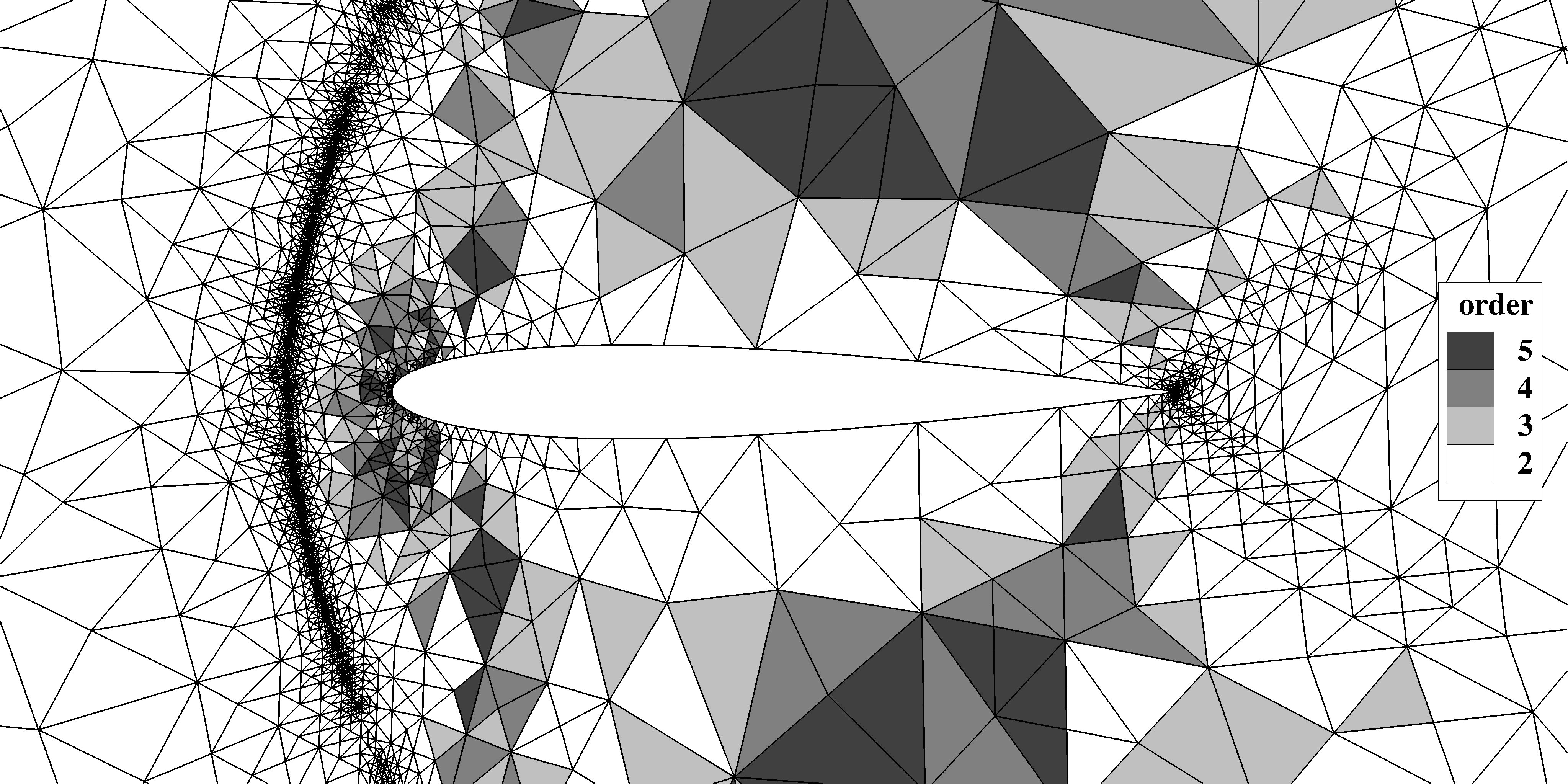}}
\caption{Adapted meshes for the supersonic Euler test case (${\rm Ma}_{\infty}=1.4$, $\alpha=0^{\circ}$)}
\end{figure}

A purely $h$-adapted mesh is given in Fig.~\ref{fig:naca_ma14_al0_mesh_h}. Several refinement regions can be identified: the bow shock, leading and trailing edge, and a feature in the adjoint solution emanating from the trailing edge in the upstream direction. During $hp$-adaptation, the latter feature and the leading edge are being treated with $p$-enrichment.

\begin{figure}[h]
  \subfloat[HDG]{\includegraphics[width=0.5\columnwidth]{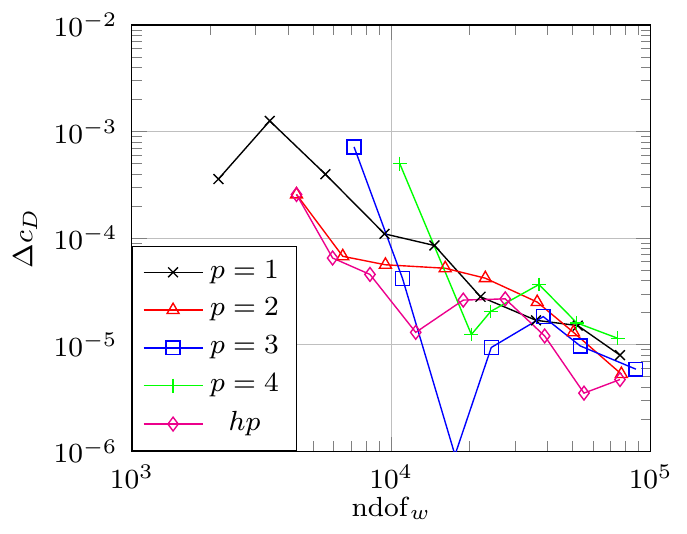}}
  \subfloat[DG]{\includegraphics[width=0.5\columnwidth]{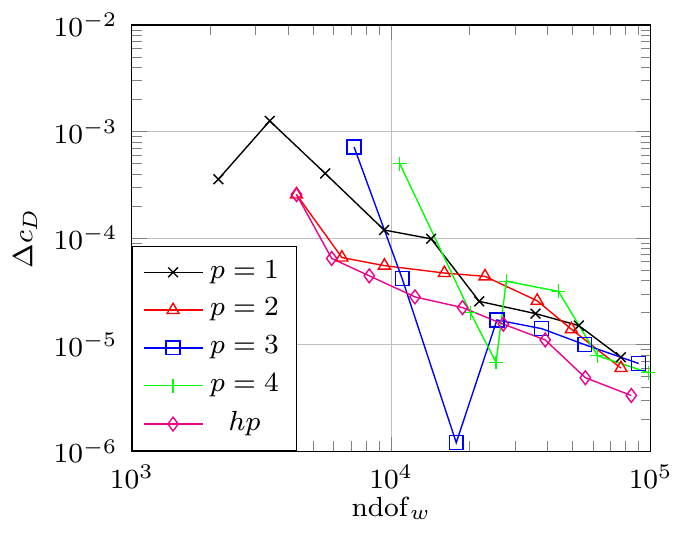}}
  \caption{Drag convergence with respect to degrees of freedom (${\rm Ma}_{\infty}=1.4$, $\alpha=0^{\circ}$)}
  \label{fig:ma14al0_ndof}  
\end{figure}

\begin{figure}[h]
  \subfloat[HDG]{\includegraphics[width=0.5\columnwidth]{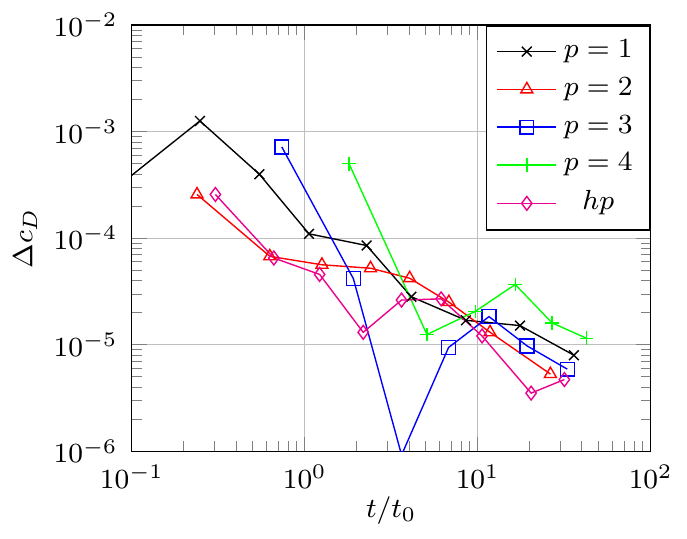}}
  \subfloat[DG]{\includegraphics[width=0.5\columnwidth]{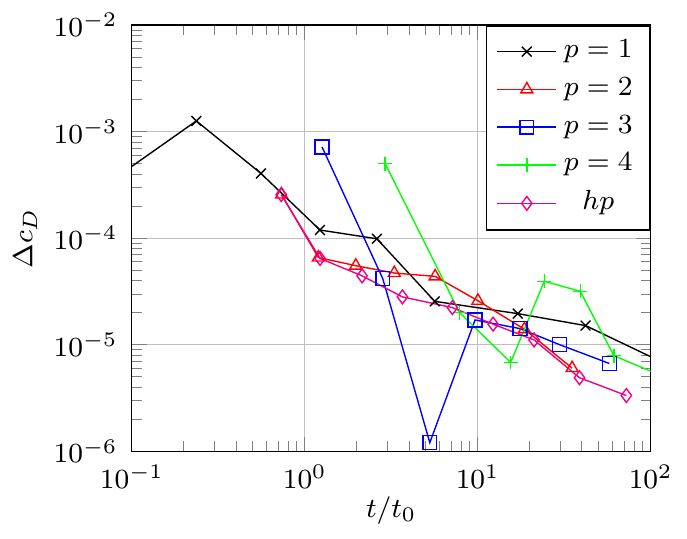}}
  \caption{Drag convergence with respect to time (${\rm Ma}_{\infty}=1.4$, $\alpha=0^{\circ}$)}
  \label{fig:ma14al0_time}  
\end{figure}

Both HDG and DG show a similar behavior in terms of error reduction with respect to degrees of freedom (see Fig.~\ref{fig:ma14al0_ndof}). A comparison between different polynomial orders is rather inconclusive which might be due to the lack of smoothness of this test case. Furthermore, we  note that the convergence behavior is not monotonic (see for example $p=3$ in Fig.~\ref{fig:ma14al0_ndof}). The error reduction with respect to computational time is similar (see Fig.~\ref{fig:ma14al0_time}). In both measures, the $hp$-adaptive run gives the overall best solution.

Considering the time to solution, HDG is approximately 50\% faster than DG (see Tbl.~\ref{tbl:runtime_supersonic}). The $hp$-adaptive run is more than 3.5 times faster. Similar to the transonic test case, the ratio of necessary nonzero entries attains its highest value for $p=4$. In the case of $hp$-adaptation, this ratio is not as high, as in the shock region a lot of elements with $p=2$ exist.

\begin{table}[h]
\centering
\begin{tabular}{c|c|c|c|c|c}
\hline 
$p$ & $1$ & $2$ & $3$ & $4$ & $hp$\\ \hline
$t_{\rm DG}/t_{\rm HDG}$ & 2.817 & 1.335 & 1.751 & 1.441 & 3.554 \\
$n_{\rm nz, DG}/n_{\rm nz, HDG}$ & 1.212 & 2.132 & 3.429 & 4.030 & 3.806 \\
\hline 
\end{tabular}
\caption{Runtime and nonzero ratios for a fixed error level ($\rm Ma_{\infty}=1.4$, $\alpha=0^{\circ}$)}
\label{tbl:runtime_supersonic}
\end{table}

\subsection{Subsonic Laminar Flow over the NACA~0012 Airfoil}

Finally, we consider viscous flow in the subsonic regime. The free stream Mach number is $\rm Ma_{\infty}=0.5$, the angle of attack $\alpha=1^{\circ}$ and the Reynolds number $\rm Re=5000$. Due to the latter, a thin boundary layer develops around the airfoil.

\begin{figure}
\centering
\includegraphics[width=\columnwidth]{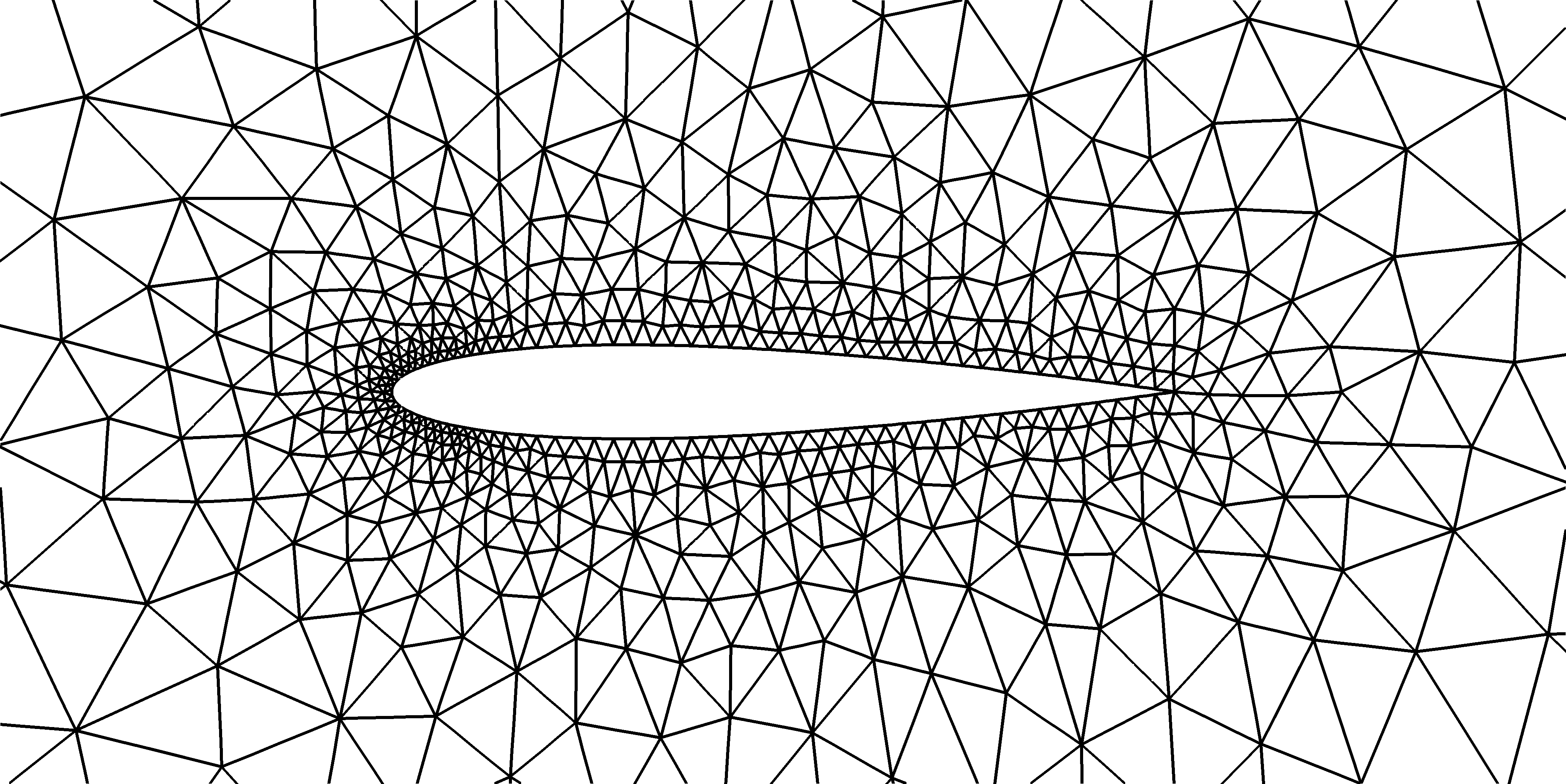}
\caption{Baseline mesh with 1781 elements for viscous computations}
\label{fig:ns_mesh_coarse}
\end{figure}

The baseline mesh for the Navier-Stokes test case is more refined around the airfoil such that the boundary layer is correctly captured (see Fig.~\ref{fig:ns_mesh_coarse}). It consists of 1781 elements and its far field is over a 1000 chords away.

Admissible target functionals defined on the boundary for the Navier-Stokes equations are given by the weighted boundary flux (including pressure and shear forces) along wall boundaries, i.e.
\begin{equation}
J\left(w,\boldsymbol\nabla w\right)=\int_{\partial\Omega}\boldsymbol\psi\cdot\left(p\mathbf{n}-\boldsymbol\tau \mathbf{n}\right)\ds
\end{equation}
where $\mathbf{n}$ is the outward pointing normal. Here, $\boldsymbol\psi$ is nonzero only on wall boundaries. By using $\boldsymbol\psi=\frac1{C_{\infty}}\left(\cos{\alpha}, \sin{\alpha}\right)^T$ or $\boldsymbol\psi=\frac1{C_{\infty}}\left(-\sin{\alpha}, \cos{\alpha}\right)^T$ along wall boundaries and otherwise 0, the functional represents the viscous drag coefficient $c_D$ or the viscous lift coefficient $c_L$, respectively.

\begin{figure}
\centering
\subfloat[Pure $h$-adaptation ($p=2$)\label{fig:naca_ma05_al1_re5000_mesh_h}]{
\includegraphics[width=\columnwidth]{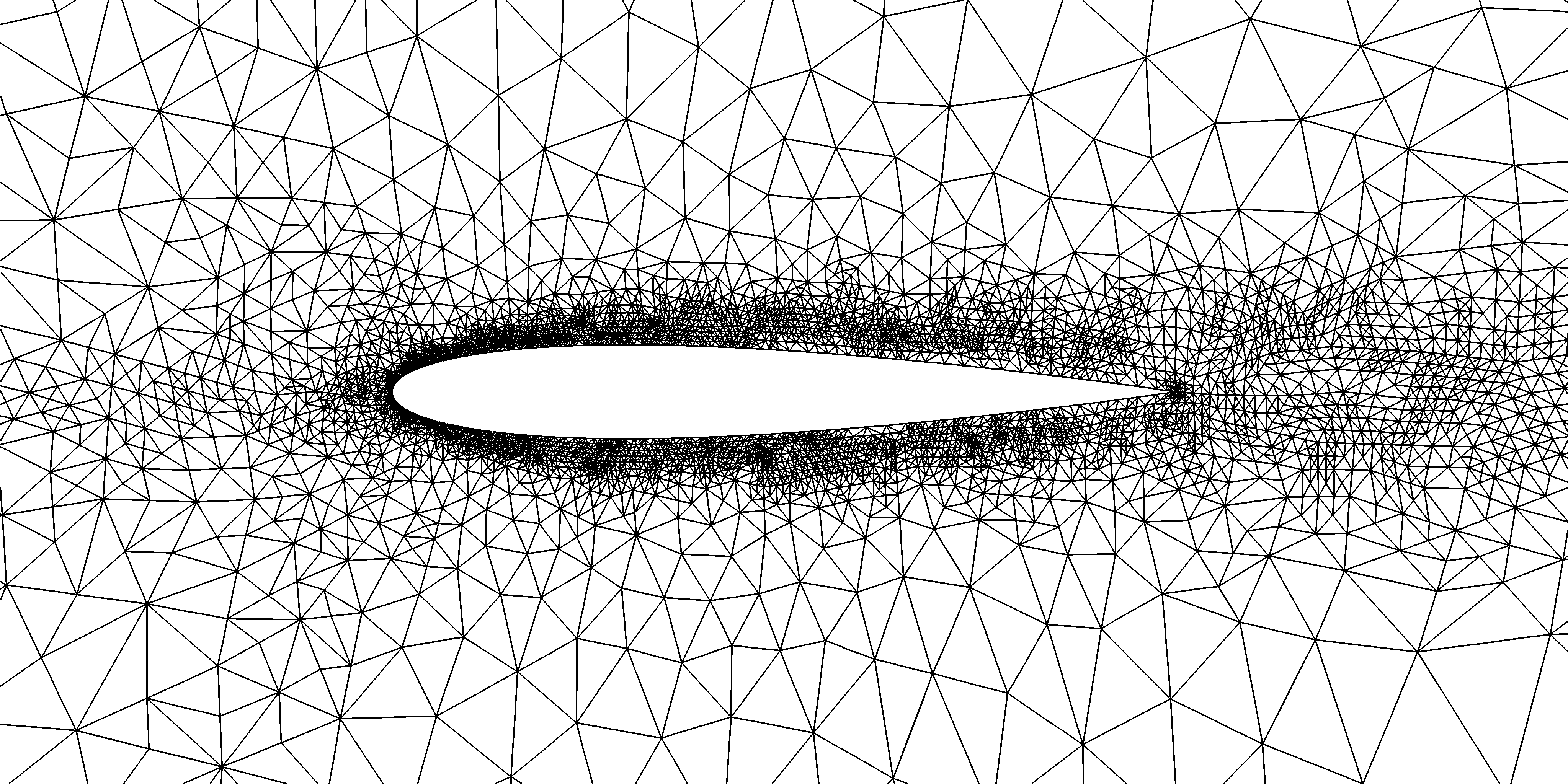}}\\
\subfloat[$hp$-adaptation ($p=2\dots 5$)\label{fig:naca_ma05_al1_re5000_mesh_hp}]{
\includegraphics[width=\columnwidth]{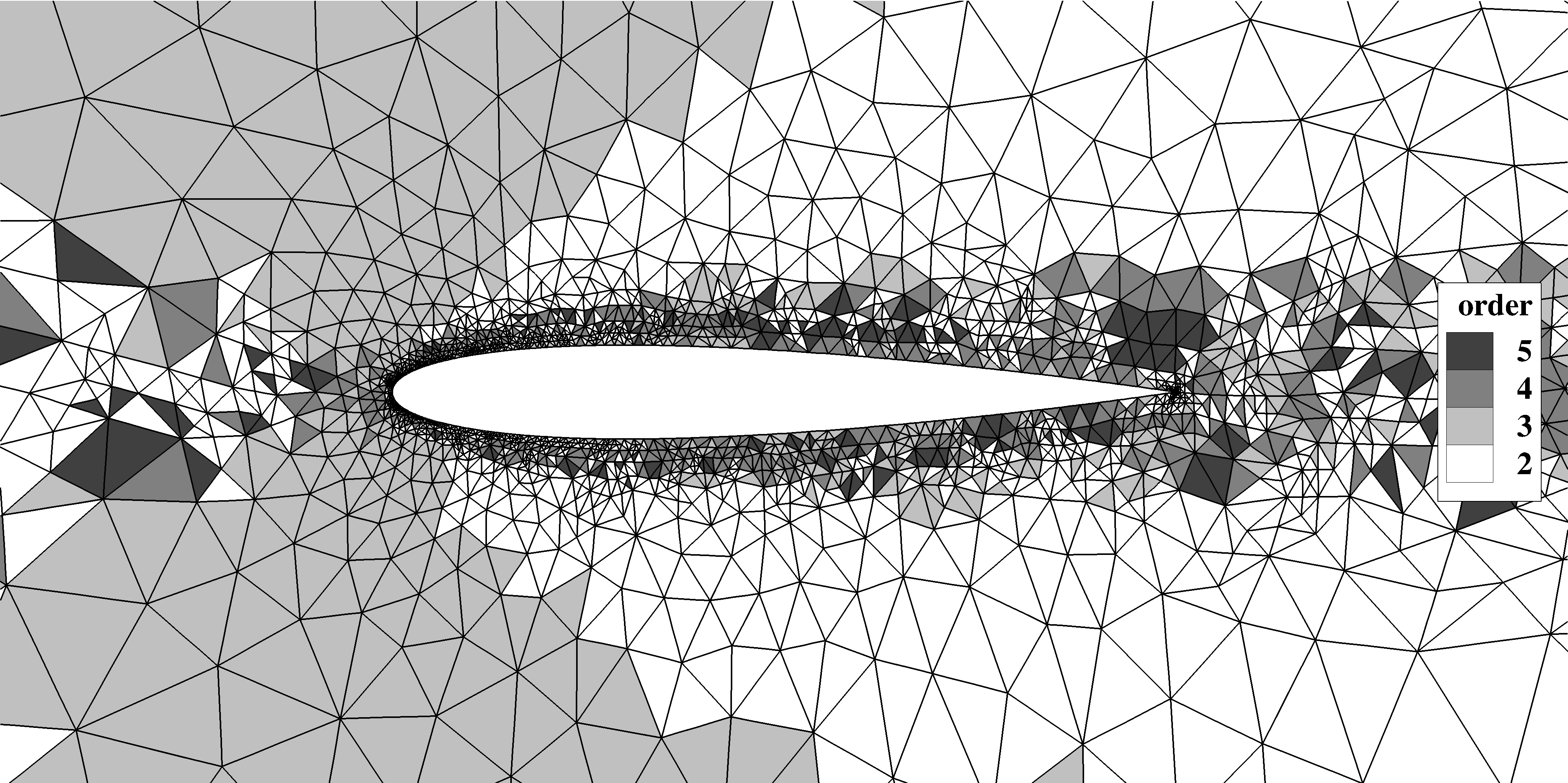}}
\caption{Adapted meshes for the Navier-Stokes test case (${\rm Ma}_{\infty}=0.5$, $\alpha=1^{\circ}$, $\rm Re=5000$)}
\end{figure}

Both the $h$-adapted mesh (see Fig.~\ref{fig:naca_ma05_al1_re5000_mesh_h}) and the $hp$-adapted mesh (see Fig.~\ref{fig:naca_ma05_al1_re5000_mesh_hp}) undergo refinement within the boundary layer and the wake region. The mesh refinement for the $hp$-adaptive run is however more restricted to the leading edge region where the boundary layer develops. Further downstream, $p$-enrichment is used as soon as the necessary mesh-resolution is reached.

\begin{figure}[h]
  \subfloat[HDG]{\includegraphics[width=0.5\columnwidth]{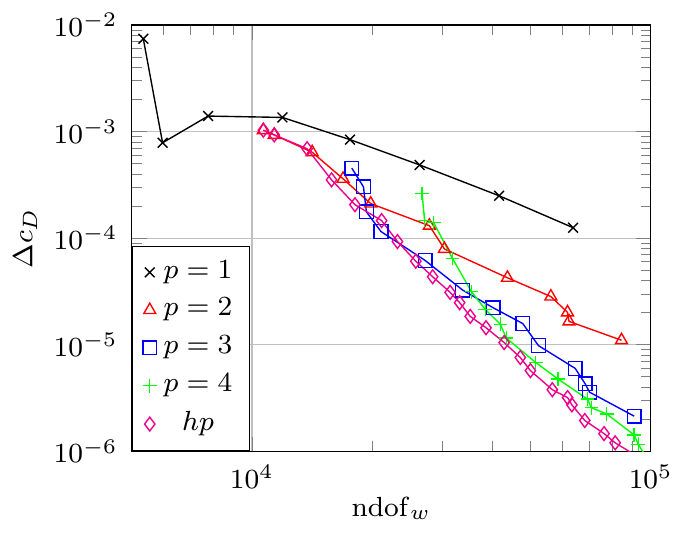}}
  \subfloat[DG]{\includegraphics[width=0.5\columnwidth]{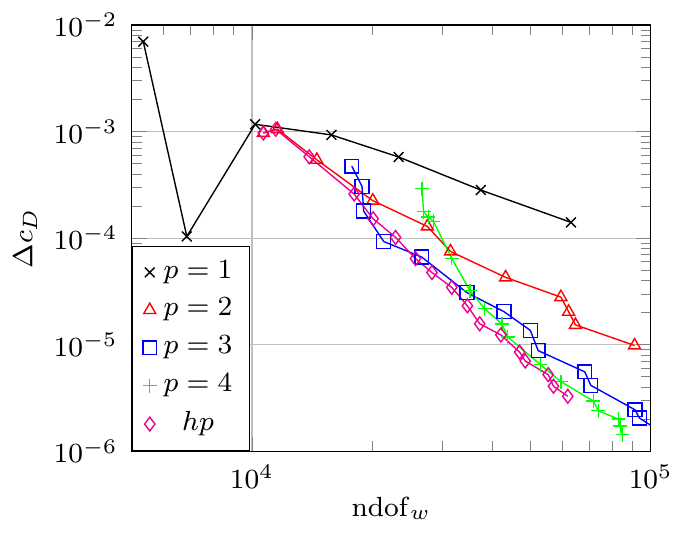}}
  \caption{Drag convergence with respect to degrees of freedom (${\rm Ma}_{\infty}=0.5$, $\alpha=1^{\circ}$, ${\rm Re}=5000$)}
  \label{fig:ma05al1re5000_ndof}  
\end{figure}

\begin{figure}[h]
  \subfloat[HDG]{\includegraphics[width=0.5\columnwidth]{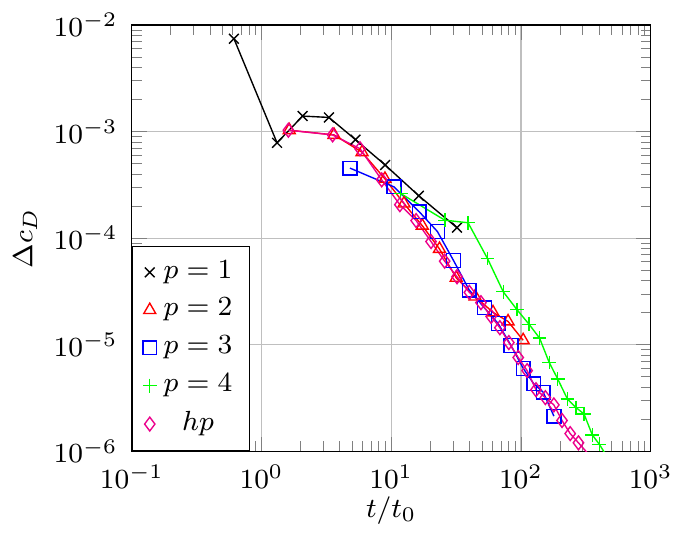}}
  \subfloat[DG]{\includegraphics[width=0.5\columnwidth]{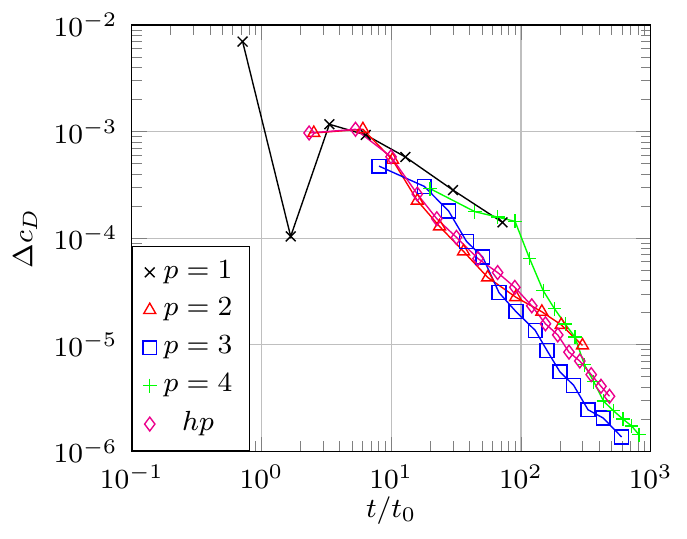}}
  \caption{Drag convergence with respect to time (${\rm Ma}_{\infty}=0.5$, $\alpha=1^{\circ}$, ${\rm Re}=5000$)}
  \label{fig:ma05al1re5000_time}    
\end{figure}

In terms of accuracy versus degrees of freedom, HDG and DG perform comparably well. The higher the polynomial degree the more accurate and efficient the computations are for both HDG and DG (see Fig.~\ref{fig:ma05al1re5000_ndof} and \ref{fig:ma05al1re5000_time}). The difference between $hp$, $p=3$ and $p=4$ is not as big, though. This might lead to the conclusion that isotropic mesh refinement is not longer efficiently applicable in cases involving strong gradients. Anisotropic refinement might be able to remedy this problem so that the full potential of $hp$-adaptation can be attained.

Concerning the timings, we can see a similar trend as in the previous test cases (see Tbl.~\ref{tbl:runtime_laminar}). 
For $p=1\dots 4$, HDG is more than twice as fast. The $hp$-adaptive HDG computation is even three times as fast compared to the DG run. For $p\geq 2$ the savings in nonzero entries for HDG become significant.

\begin{table}
\centering
\begin{tabular}{c|c|c|c|c|c}
\hline 
$p$ & $1$ & $2$ & $3$ & $4$ & $hp$\\ \hline
$t_{\rm DG}/t_{\rm HDG}$ & 2.244 & 2.572 & 2.397 & 2.288 & 3.122 \\
$n_{\rm nz, DG}/n_{\rm nz, HDG}$ & 1.196 & 2.235 & 3.472 & 4.531 & 3.463 \\
\hline
\end{tabular}
\caption{Runtime and nonzero ratios for a fixed error level ($\rm Ma_{\infty}=0.5$, $\alpha=1^{\circ}$, $\rm Re=5000$)}
\label{tbl:runtime_laminar}
\end{table}

%% file: conclusion.tex

\section{Conclusion and Outlook}
\label{sec:conclusions}
We presented an adjoint-based $hp$-adaptation methodology and compared it for hybridized and non-hybridized discontinuous Galerkin methods. $hp$-adaptation proved to be superior to pure $h$-adaptation if discontinuous or singular flow features were involved. In all cases, a higher polynomial degree turned out to be beneficial. We showed that one can expect HDG to be better than DG in terms of runtime and memory requirements from $p=2$ on for a broad range of test cases. The results are promising enough to pursue further comparisons with more complex test cases.

We plan to extend our computational framework to three dimensional problems \cite{woopen2014hdg3d}. Then, adaptivity will play an even more crucial role, as the problem size increases drastically compared to the two dimensional case.

Furthermore, compressible flows are often dominated by anisotropic features, such as shocks or very thin boundary layers. Thus, taking this anisotropy into account during adaptation is crucial. We plan to incorporate anisotropic adaptation within future projects. 

%% file: main.bbl
\begin{thebibliography}{34}
\providecommand{\natexlab}[1]{#1}
\providecommand{\url}[1]{\texttt{#1}}
\providecommand{\urlprefix}{URL }
\expandafter\ifx\csname urlstyle\endcsname\relax
  \providecommand{\doi}[1]{doi:\discretionary{}{}{}#1}\else
  \providecommand{\doi}[1]{doi:\discretionary{}{}{}\begingroup
  \urlstyle{rm}\url{#1}\endgroup}\fi
\providecommand{\bibinfo}[2]{#2}

\bibitem[{Bassi and Rebay(1997)}]{Bassi1997}
\bibinfo{author}{F.~Bassi}, \bibinfo{author}{S.~Rebay}, \bibinfo{title}{{A
  High-Order Accurate Discontinuous {Finite-Element} Method for the Numerical
  Solution of the Compressible {Navier-Stokes} Equations}},
  \bibinfo{journal}{Journal of Computational Physics} \bibinfo{volume}{131}
  (\bibinfo{year}{1997}) \bibinfo{pages}{267--279}.

\bibitem[{Cockburn and Shu(1988)}]{DG2}
\bibinfo{author}{B.~Cockburn}, \bibinfo{author}{C.-W. Shu},
  \bibinfo{title}{{TVB Runge-Kutta} Local Projection {Discontinuous Galerkin}
  Finite Element Method for {Conservation Laws} {II}: General Framework},
  \bibinfo{journal}{Mathematics of Computation} \bibinfo{volume}{52}
  (\bibinfo{year}{1988}) \bibinfo{pages}{411--435}.

\bibitem[{Arnold et~al.(2002)Arnold, Brezzi, Cockburn, and Marini}]{ABCM}
\bibinfo{author}{D.~Arnold}, \bibinfo{author}{F.~Brezzi},
  \bibinfo{author}{B.~Cockburn}, \bibinfo{author}{L.~Marini},
  \bibinfo{title}{{Unified Analysis of Discontinuous Galerkin Methods for
  Elliptic Problems}}, \bibinfo{journal}{SIAM Journal on Numerical Analysis}
  \bibinfo{volume}{39} (\bibinfo{year}{2002}) \bibinfo{pages}{1749--1779}.

\bibitem[{de~Veubeke(1965)}]{fraeijs1965displacement}
\bibinfo{author}{B.~F. de~Veubeke}, \bibinfo{title}{{Displacement and
  Equilibrium Models in the Finite Element Method}},
  chap.~\bibinfo{chapter}{9}, \bibinfo{publisher}{Wiley New York},
  \bibinfo{pages}{145--197}, \bibinfo{year}{1965}.

\bibitem[{Cockburn and Gopalakrishnan(2004)}]{cockburn2004characterization}
\bibinfo{author}{B.~Cockburn}, \bibinfo{author}{J.~Gopalakrishnan},
  \bibinfo{title}{{A Characterization of Hybridized Mixed Methods for Second
  Order Elliptic Problems}}, \bibinfo{journal}{SIAM Journal on Numerical
  Analysis} \bibinfo{volume}{42}~(\bibinfo{number}{1}) (\bibinfo{year}{2004})
  \bibinfo{pages}{283--301}.

\bibitem[{Cockburn et~al.(2009)Cockburn, Gopalakrishnan, and
  Lazarov}]{cockburn2009unified}
\bibinfo{author}{B.~Cockburn}, \bibinfo{author}{J.~Gopalakrishnan},
  \bibinfo{author}{R.~Lazarov}, \bibinfo{title}{{Unified Hybridization of
  Discontinuous Galerkin, Mixed, and Continuous Galerkin Methods for Second
  Order Elliptic Problems}}, \bibinfo{journal}{SIAM Journal on Numerical
  Analysis} \bibinfo{volume}{47}~(\bibinfo{number}{2}) (\bibinfo{year}{2009})
  \bibinfo{pages}{1319--1365}.

\bibitem[{Nguyen et~al.(2009{\natexlab{a}})Nguyen, Peraire, and
  Cockburn}]{nguyen2009implicitlin}
\bibinfo{author}{N.~Nguyen}, \bibinfo{author}{J.~Peraire},
  \bibinfo{author}{B.~Cockburn}, \bibinfo{title}{{An Implicit High-Order
  Hybridizable Discontinuous Galerkin Method for Linear Convection-Diffusion
  Equations}}, \bibinfo{journal}{Journal of Computational Physics}
  \bibinfo{volume}{228}~(\bibinfo{number}{9})
  (\bibinfo{year}{2009}{\natexlab{a}}) \bibinfo{pages}{3232--3254}.

\bibitem[{Nguyen et~al.(2009{\natexlab{b}})Nguyen, Peraire, and
  Cockburn}]{NPC09}
\bibinfo{author}{N.~Nguyen}, \bibinfo{author}{J.~Peraire},
  \bibinfo{author}{B.~Cockburn}, \bibinfo{title}{{An Implicit High-Order
  Hybridizable Discontinuous Galerkin Method for Nonlinear Convection-Diffusion
  Equations}}, \bibinfo{journal}{Journal of Computational Physics}
  \bibinfo{volume}{228} (\bibinfo{year}{2009}{\natexlab{b}})
  \bibinfo{pages}{8841--8855}.

\bibitem[{Peraire et~al.(2010)Peraire, Nguyen, and
  Cockburn}]{peraire2010hybridizable}
\bibinfo{author}{J.~Peraire}, \bibinfo{author}{N.~Nguyen},
  \bibinfo{author}{B.~Cockburn}, \bibinfo{title}{{A Hybridizable Discontinuous
  Galerkin Method for the Compressible Euler and Navier-Stokes Equations}},
  \bibinfo{number}{AIAA Paper 2010-0363}, \bibinfo{year}{2010}.

\bibitem[{Huerta et~al.(2013)Huerta, Angeloski, Roca, and Peraire}]{AA-HARP:13}
\bibinfo{author}{A.~Huerta}, \bibinfo{author}{A.~Angeloski},
  \bibinfo{author}{X.~Roca}, \bibinfo{author}{J.~Peraire},
  \bibinfo{title}{{Efficiency of High-Order Elements for Continuous and
  Discontinuous Galerkin Methods}}, \bibinfo{journal}{Int. J. Numer. Methods
  Eng.} \bibinfo{volume}{96}~(\bibinfo{number}{9}) (\bibinfo{year}{2013})
  \bibinfo{pages}{529--560}.

\bibitem[{Hartmann and Houston(2002)}]{hartmann2002adaptive}
\bibinfo{author}{R.~Hartmann}, \bibinfo{author}{P.~Houston},
  \bibinfo{title}{{Adaptive Discontinuous Galerkin Finite Element Methods for
  the Compressible Euler Equations}}, \bibinfo{journal}{Journal of
  Computational Physics} \bibinfo{volume}{183}~(\bibinfo{number}{2})
  (\bibinfo{year}{2002}) \bibinfo{pages}{508--532}.

\bibitem[{Venditti and Darmofal(2002)}]{venditti2002grid}
\bibinfo{author}{D.~Venditti}, \bibinfo{author}{D.~Darmofal},
  \bibinfo{title}{{Grid Adaptation for Functional Outputs: Application to
  Two-Dimensional Inviscid Flows}}, \bibinfo{journal}{Journal of Computational
  Physics} \bibinfo{volume}{176}~(\bibinfo{number}{1}) (\bibinfo{year}{2002})
  \bibinfo{pages}{40--69}.

\bibitem[{Venditti and Darmofal(2003)}]{venditti2003anisotropic}
\bibinfo{author}{D.~Venditti}, \bibinfo{author}{D.~Darmofal},
  \bibinfo{title}{{Anisotropic Grid Adaptation for Functional Outputs:
  Application to Two-Dimensional Viscous Flows}}, \bibinfo{journal}{Journal of
  Computational Physics} \bibinfo{volume}{187}~(\bibinfo{number}{1})
  (\bibinfo{year}{2003}) \bibinfo{pages}{22--46}.

\bibitem[{Hartmann(2006)}]{hartmann2006adaptive}
\bibinfo{author}{R.~Hartmann}, \bibinfo{title}{{Adaptive Discontinuous Galerkin
  Methods with Shock-Capturing for the Compressible Navier--Stokes Equations}},
  \bibinfo{journal}{International Journal for Numerical Methods in Fluids}
  \bibinfo{volume}{51}~(\bibinfo{number}{9-10}) (\bibinfo{year}{2006})
  \bibinfo{pages}{1131--1156}.

\bibitem[{Fidkowski and Darmofal(2011)}]{fidkowski2011review}
\bibinfo{author}{K.~Fidkowski}, \bibinfo{author}{D.~Darmofal},
  \bibinfo{title}{{Review of Output-Based Error Estimation and Mesh Adaptation
  in Computational Fluid Dynamics}}, \bibinfo{journal}{AIAA Journal}
  \bibinfo{volume}{49}~(\bibinfo{number}{4}) (\bibinfo{year}{2011})
  \bibinfo{pages}{673--694}.

\bibitem[{Giorgiani et~al.(2013)Giorgiani, Fern{\'a}ndez-M{\'e}ndez, and
  Huerta}]{giorgiani2013hybridizable}
\bibinfo{author}{G.~Giorgiani}, \bibinfo{author}{S.~Fern{\'a}ndez-M{\'e}ndez},
  \bibinfo{author}{A.~Huerta}, \bibinfo{title}{{Hybridizable Discontinuous
  Galerkin $p$-Adaptivity for Wave Propagation Problems}},
  \bibinfo{journal}{International Journal for Numerical Methods in Fluids} .

\bibitem[{Sch{\"u}tz and May(2013{\natexlab{a}})}]{schutz201358}
\bibinfo{author}{J.~Sch{\"u}tz}, \bibinfo{author}{G.~May}, \bibinfo{title}{{A
  Hybrid Mixed Method for the Compressible Navier--Stokes Equations}},
  \bibinfo{journal}{Journal of Computational Physics} \bibinfo{volume}{240}
  (\bibinfo{year}{2013}{\natexlab{a}}) \bibinfo{pages}{58--75}.

\bibitem[{Sch{\"u}tz and May(2013{\natexlab{b}})}]{schutz2012adjoint}
\bibinfo{author}{J.~Sch{\"u}tz}, \bibinfo{author}{G.~May}, \bibinfo{title}{{An
  Adjoint Consistency Analysis for a Class of Hybrid Mixed Methods}},
  \bibinfo{journal}{IMA Journal of Numerical Analysis} .

\bibitem[{Woopen et~al.(2013)Woopen, May, and Sch{\"u}tz}]{woopen2013adjoint}
\bibinfo{author}{M.~Woopen}, \bibinfo{author}{G.~May},
  \bibinfo{author}{J.~Sch{\"u}tz}, \bibinfo{title}{{Adjoint-Based Error
  Estimation and Mesh Adaptation for Hybridized Discontinuous Galerkin
  Methods}}, \bibinfo{journal}{arXiv preprint arXiv:1309.3649} .

\bibitem[{Balan et~al.(2013)Balan, Woopen, and May}]{balan2013hybridadjointhp}
\bibinfo{author}{A.~Balan}, \bibinfo{author}{M.~Woopen},
  \bibinfo{author}{G.~May}, \bibinfo{title}{{Adjoint-Based $hp$-Adaptation for
  a Class of High-Order Hybridized Finite Element Schemes for Compressible
  Flows}}, \bibinfo{number}{AIAA Paper 2013-2938}, \bibinfo{year}{2013}.

\bibitem[{Godlewski and Raviart(1996)}]{godlewski1996}
\bibinfo{author}{E.~Godlewski}, \bibinfo{author}{P.~Raviart},
  \bibinfo{title}{{Numerical Approximation of Hyperbolic Systems of
  Conservation Laws}}, no. \bibinfo{number}{Nr. 118} in
  \bibinfo{series}{Applied Mathematical Sciences},
  \bibinfo{publisher}{Springer}, ISBN \bibinfo{isbn}{9780387945293},
  \bibinfo{year}{1996}.

\bibitem[{Chen and Cockburn(2012)}]{chen2012analysis}
\bibinfo{author}{Y.~Chen}, \bibinfo{author}{B.~Cockburn},
  \bibinfo{title}{{Analysis of Variable-Degree HDG Methods for
  Convection--Diffusion Equations. Part I: General Nonconforming Meshes}},
  \bibinfo{journal}{IMA Journal of Numerical Analysis}
  \bibinfo{volume}{32}~(\bibinfo{number}{4}) (\bibinfo{year}{2012})
  \bibinfo{pages}{1267--1293}.

\bibitem[{Cockburn and Gopalakrishnan(2005)}]{cockburn2005error}
\bibinfo{author}{B.~Cockburn}, \bibinfo{author}{J.~Gopalakrishnan},
  \bibinfo{title}{{Error Analysis of Variable Degree Mixed Methods for Elliptic
  Problems via Hybridization}}, \bibinfo{journal}{Mathematics of Computation}
  \bibinfo{volume}{74}~(\bibinfo{number}{252}) (\bibinfo{year}{2005})
  \bibinfo{pages}{1653--1677}.

\bibitem[{Mavriplis and Jameson(1990)}]{mavriplis1990multigrid}
\bibinfo{author}{D.~Mavriplis}, \bibinfo{author}{A.~Jameson},
  \bibinfo{title}{{Multigrid Solution of the Navier-Stokes Equations on
  Triangular Meshes}}, \bibinfo{journal}{AIAA Journal}
  \bibinfo{volume}{28}~(\bibinfo{number}{8}) (\bibinfo{year}{1990})
  \bibinfo{pages}{1415--1425}.

\bibitem[{Balay et~al.(2013{\natexlab{a}})Balay, Brown, Buschelman, Gropp,
  Kaushik, Knepley, McInnes, Smith, and Zhang}]{petsc-web-page}
\bibinfo{author}{S.~Balay}, \bibinfo{author}{J.~Brown},
  \bibinfo{author}{K.~Buschelman}, \bibinfo{author}{W.~Gropp},
  \bibinfo{author}{D.~Kaushik}, \bibinfo{author}{M.~Knepley},
  \bibinfo{author}{L.~McInnes}, \bibinfo{author}{B.~Smith},
  \bibinfo{author}{H.~Zhang}, \bibinfo{title}{{PETSc Web Page}},
  \bibinfo{note}{http://www.mcs.anl.gov/petsc},
  \bibinfo{year}{2013}{\natexlab{a}}.

\bibitem[{Balay et~al.(2013{\natexlab{b}})Balay, Brown, Buschelman, Eijkhout,
  Gropp, Kaushik, Knepley, McInnes, Smith, and Zhang}]{petsc-user-ref}
\bibinfo{author}{S.~Balay}, \bibinfo{author}{J.~Brown},
  \bibinfo{author}{K.~Buschelman}, \bibinfo{author}{V.~Eijkhout},
  \bibinfo{author}{W.~Gropp}, \bibinfo{author}{D.~Kaushik},
  \bibinfo{author}{M.~Knepley}, \bibinfo{author}{L.~McInnes},
  \bibinfo{author}{B.~Smith}, \bibinfo{author}{H.~Zhang},
  \bibinfo{title}{{PETSc Users Manual}}, \bibinfo{type}{Tech. Rep.}
  \bibinfo{number}{ANL-95/11 - Revision 3.4}, \bibinfo{institution}{Argonne
  National Laboratory}, \bibinfo{year}{2013}{\natexlab{b}}.

\bibitem[{D{\"o}rfler(1996)}]{dorfler1996convergent}
\bibinfo{author}{W.~D{\"o}rfler}, \bibinfo{title}{{A Convergent Adaptive
  Algorithm for Poisson's Equation}}, \bibinfo{journal}{SIAM Journal on
  Numerical Analysis} \bibinfo{volume}{33}~(\bibinfo{number}{3})
  (\bibinfo{year}{1996}) \bibinfo{pages}{1106--1124}.

\bibitem[{Ceze and Fidkowski(2010)}]{ceze2010output}
\bibinfo{author}{M.~Ceze}, \bibinfo{author}{K.~Fidkowski},
  \bibinfo{title}{{Output-Driven Anisotropic Mesh Adaptation for Viscous Flows
  using Discrete Choice Optimization}}, \bibinfo{number}{AIAA Paper 2010-0170},
  \bibinfo{year}{2010}.

\bibitem[{Wang and Mavriplis(2009)}]{wangjcp2009}
\bibinfo{author}{L.~Wang}, \bibinfo{author}{D.~J. Mavriplis},
  \bibinfo{title}{{Adjoint-Based $h$-$p$ Adaptive Discontinuous Galerkin
  Methods for the 2D Compressible Euler Equations}}, \bibinfo{journal}{Journal
  of Computational Physics} \bibinfo{volume}{228} (\bibinfo{year}{2009})
  \bibinfo{pages}{7643--7661}.

\bibitem[{Persson and Peraire(2006)}]{persson2006shockcapturing}
\bibinfo{author}{P.~Persson}, \bibinfo{author}{J.~Peraire},
  \bibinfo{title}{{Sub-cell Shock Capturing for Discontinuous Galerkin
  Methods}}, \bibinfo{number}{AIAA Paper 2006-0112}, \bibinfo{year}{2006}.

\bibitem[{Dubiner(1991)}]{dubiner:1991triagTensorBasis}
\bibinfo{author}{M.~Dubiner}, \bibinfo{title}{Spectral Methods on Triangles and
  Other Domains}, \bibinfo{journal}{J. Sci. Comput.}
  \bibinfo{volume}{6}~(\bibinfo{number}{4}) (\bibinfo{year}{1991})
  \bibinfo{pages}{345--390}.

\bibitem[{Bassi et~al.(2005)Bassi, Crivellini, Rebay, and
  Savini}]{bassi2005discontinuous}
\bibinfo{author}{F.~Bassi}, \bibinfo{author}{A.~Crivellini},
  \bibinfo{author}{S.~Rebay}, \bibinfo{author}{M.~Savini},
  \bibinfo{title}{{Discontinuous Galerkin Solution of the Reynolds-Averaged
  Navier-Stokes and k-$\omega$ Turbulence Model Equations}},
  \bibinfo{journal}{Computers \& Fluids}
  \bibinfo{volume}{34}~(\bibinfo{number}{4}) (\bibinfo{year}{2005})
  \bibinfo{pages}{507--540}.

\bibitem[{Hartmann(2007)}]{hartmann2007adjoint}
\bibinfo{author}{R.~Hartmann}, \bibinfo{title}{{Adjoint Consistency Analysis of
  Discontinuous Galerkin Discretizations}}, \bibinfo{journal}{SIAM Journal on
  Numerical Analysis} \bibinfo{volume}{45}~(\bibinfo{number}{6})
  (\bibinfo{year}{2007}) \bibinfo{pages}{2671--2696}.

\bibitem[{Woopen et~al.(2014)Woopen, Balan, and May}]{woopen2014hdg3d}
\bibinfo{author}{M.~Woopen}, \bibinfo{author}{A.~Balan},
  \bibinfo{author}{G.~May}, \bibinfo{title}{{A Hybridized Discontinuous
  Galerkin Method for Three-Dimensional Compressible Flow Problems}},
  \bibinfo{number}{AIAA Paper 2014-0938}, \bibinfo{year}{2014}.

\end{thebibliography}
